\documentclass[aps,prx,twocolumn,amsmath,nofootinbib,longbibliography,amssymb,superscriptaddress,10pt]{revtex4-2}
\usepackage[english]{babel}
\usepackage{graphicx}
\usepackage{ulem}
\usepackage{soul}
\usepackage{float} %added by Evyatar
\usepackage[svgnames]{xcolor} %added by Evyatar
\usepackage[colorlinks, linkcolor=Blue , citecolor= Blue,urlcolor = NavyBlue, breaklinks=true]{hyperref}
\usepackage{verbatim}
\usepackage{esint}
\usepackage{marginnote}
\usepackage{graphicx,mathtools,bm,bbm}
\usepackage{multirow}
\usepackage{amsmath} % or simply amstext

\usepackage{tikz-qtree}
\usepackage{tikz}

\newcommand{\RNum}[1]{\uppercase\expandafter{\romannumeral #1\relax}}

\def \beq {\begin{eqnarray}}
\def \eeq {\end{eqnarray}}

\begin{document}

\title{Probing the order parameter symmetry of two-dimensional superconductors by twisted Josephson interferometry}
\author{Jiewen Xiao, Yaar Vituri, and Erez Berg}
\affiliation{Department of Condensed Matter Physics, Weizmann Institute of Science, Rehovot 76100, Israel}
\date{\today}

\begin{abstract}
Probing the superconducting order parameter symmetry is a crucial step towards understanding the pairing mechanism in unconventional superconductors. 
%Phase-sensitive techniques, such as superconducting quantum interference device (SQUID) and Josephson junctions, are strong tools to determine the order parameter phase, which were successfully applied to the $d$-wave high Tc cuprates. However, given various superconductors in different materials, the order parameter symmetry in most of them remains unknown, especially for the recently discovered two-dimensional van der Waals superconductors, for example, magic angle twisted bilayer graphene (MATBG). 
%The superconductivity in these systems is a rich interplay between flat isolated bands, strong interactions and multi-flavor symmetry. However, order parameter remains unknown so far. 
Inspired by the recent discoveries of superconductivity in various van der Waals materials, and the availability of the relative twist angle as a continuous tuning knob in these systems, 
%Inspired by the rotation degrees of freedom, 
%here 
we propose a general setup for probing the order parameter symmetry of two-dimensional superconductors in twisted Josephson junctions. The junction is composed of an anisotropic $s$-wave superconductor as a probe and another superconductor with an unknown order parameter symmetry.  
Assuming momentum-resolved tunneling, we investigate signatures of different order parameter symmetries in the twist angle dependence of the critical current, the current-phase relations, and magnetic field dependence.
As a concrete example, we study a twisted Josephson junction between NbSe$_2$ and magic angle twisted bilayer graphene. 
% , to demonstrate how the orbital symmetry of the order parameter in MATBG can be probed, but it is a very general method that can apply to many vdW superconductors. 
%Different signatures of the current-phase relations from different order parameters can be potentially probed by fractional Shapiro steps and Josephson diode effect. 

\end{abstract}

\maketitle

\section{Introduction}
Identifying the pairing symmetry of unconventional superconductors (SCs) is a central challenge in condensed matter physics. 
%One of the most intriguing questions in condensed matter physics is the symmetry of a superconductor (SC) order parameter and its pairing mechanism.
% , which are closely related as one follows from the other. 
% Knowing the order parameter symmetry is a crucial to understand the possible pairing mechanisms.
% For conventional $s$-wave SCs, 
% %in a large proportion of materials, 
% the celebrated Bardeen–Cooper–Schrieffer (BCS) theory gives a successful microscopic description.
% For unconventional SCs, for example, high-Tc cuprates, the order parameter has a complicated spatial structure that requires a different pairing mechanism. 
This is often the key step towards understanding the pairing mechanism. 
The structure of the order parameter may be probed by various experimental techniques, divided into non-phase-sensitive and phase-sensitive methods. 
% \textcolor{blue}{(Maybe we should emphasize that while for high Tc it is mostly known, for many VdW materials this question is still unanswered)}
Non-phase-sensitive methods probe the excitation spectrum, searching for gapless (nodal) quasi-particles.  
%thermodynamic and excited states properties related to the nodal or gapped superconducting states. 
% However, they are limited in distinguishing between different types of gapped superconducting states.
Phase-sensitive techniques, such as %superconducting quantum interference device (SQUID), Josephson junction 
Josephson interferometry  \cite{van1995phase,tsuei2000pairing},  %and quasiparticle interference (QPI) by scanning tunneling microscopy (STM) \cite{hoffman2002imaging,hanaguri2007quasiparticle}, 
are based on the interference of quantum mechanical phase of the SC order parameter. 
These methods have been successfully applied to determine the nodal $d$-wave nature of the superconducting gap in the high-T$_c$ cuprate SCs.
% These techniques are extensively applied in high temperature SCs and demonstrate its $d$ wave nature, for example, the corner SQUID interferometer design \cite{van1995phase,tsuei2000pairing}. 
% \textcolor{blue}{(Do we adress non-phase-sensitive techniques? maybe a word about DOS measurements from Yazdani's group should be added)}

However, the order parameter symmetry of numerous SCs remains unknown. 
%lots of materials exhibit superconducting states at low temperatures, and their pairing mechanisms vary and remain unknown, 
Among these are the recently discovered superconducting phases in graphene multilayers,  
%, transition metal dichalcogenides (TMD), and their heterostructures. 
%For instance, superconductivity was found 
%in 
including twisted bilayer graphene (TBG) \cite{cao2018unconventional,cao2018correlated,yankowitz2019tuning}, twisted trilayer graphene (TTG) \cite{park2021tunable,cao2021pauli}, twisted structures with four and five layers \cite{park2022robust,burg2022emergence}, rhombohedral trilayer graphene (RLG) \cite{zhou2021superconductivity}, and Bernal bilayer graphene (BLG) \cite{zhou2022isospin, zhang2023enhanced}. 
% In these systems, flat band emerges and interactions energy scale becomes comparable to the kinetic energy. In addition to superconductivity, other symmetry breaking phases, for instance, correlated insulating states, also emerge in the nearby parameter regions.   
The presence of multiple electron flavors in these systems, 
%multi-flavor symmetry between 
including the valley, spin, and layer indices, gives rise to a rich phase space for electrons to pair \cite{lake2022pairing, Khalaf2022, chatterjee2022inter,PhysRevLett.127.247001}. In some of these systems, large violations of the Pauli limit have been observed~\cite{cao2021pauli,park2022robust,burg2022emergence,zhou2021superconductivity,zhou2022isospin}, indicating triplet pairing. 
% , for instance, the possible spin/valley single/triplet Cooper pairs  
% Although tremendous experiments give some evidence of the pairing nature 
In TBG and TTG, scanning tunneling microscopy (STM) experiments  \cite{oh2021evidence,kim2022evidence} have found evidence for gap nodes. Combined with the transport evidence for rotational symmetry breaking in the SC state \cite{cao2021nematicity}, these experiments indicate a non-$s$ wave pairing symmetry~\cite{lake2022pairing}.

% the order parameter symmetry remains elusive so far.

% The phase-sensitive c-axis Josephson junctions can be potentially designed to probe the order parameter symmetry, especially combining the rotation degrees of freedom. 
% The order parameter symmetry can be probed by Josephson tunneling perpendicular to the plane, especially when combined with the twis 
Twisted Josephson junctions, in which two planar superconductors are rotated relative to each other, can provide information about the pairing symmetry. 
For instance, in a twisted c-axis Josephson junction between high T$_c$ cuprates, the twist angle dependence of the critical current should reflect the $d-$wave symmetry of the order parameter~\cite{Bille2001,klemm2005phase,can2021high,volkov2021josephson,tummuru2022josephson,tummuru2022twisted,haenel2022incoherent,Song2022}.
Experimental results on such junctions have been inconsistent~\cite{li1999bi,zhao2021emergent,XueTwist}: some experiments~\cite{zhao2021emergent} have detected the predicted twisted angle dependence and others have not.
% has been studied, and the non-trivial second-order Josephson effect at $45^\circ$ twist angle reflects the $d$ wave nature \cite{li1999bi}. 
Compared to cuprates, heterostructures of 
van der Waals (vdW) materials such as graphene and transition metal dichalcogenides (TMDs) are better controlled,  
and clean interfaces exhibiting momentum resolved tunneling have been demonstrated \cite{britnell2013resonant,mishchenko2014twist,inbar2023quantum}.

\begin{figure}
    \centering
    \includegraphics[width=\linewidth]{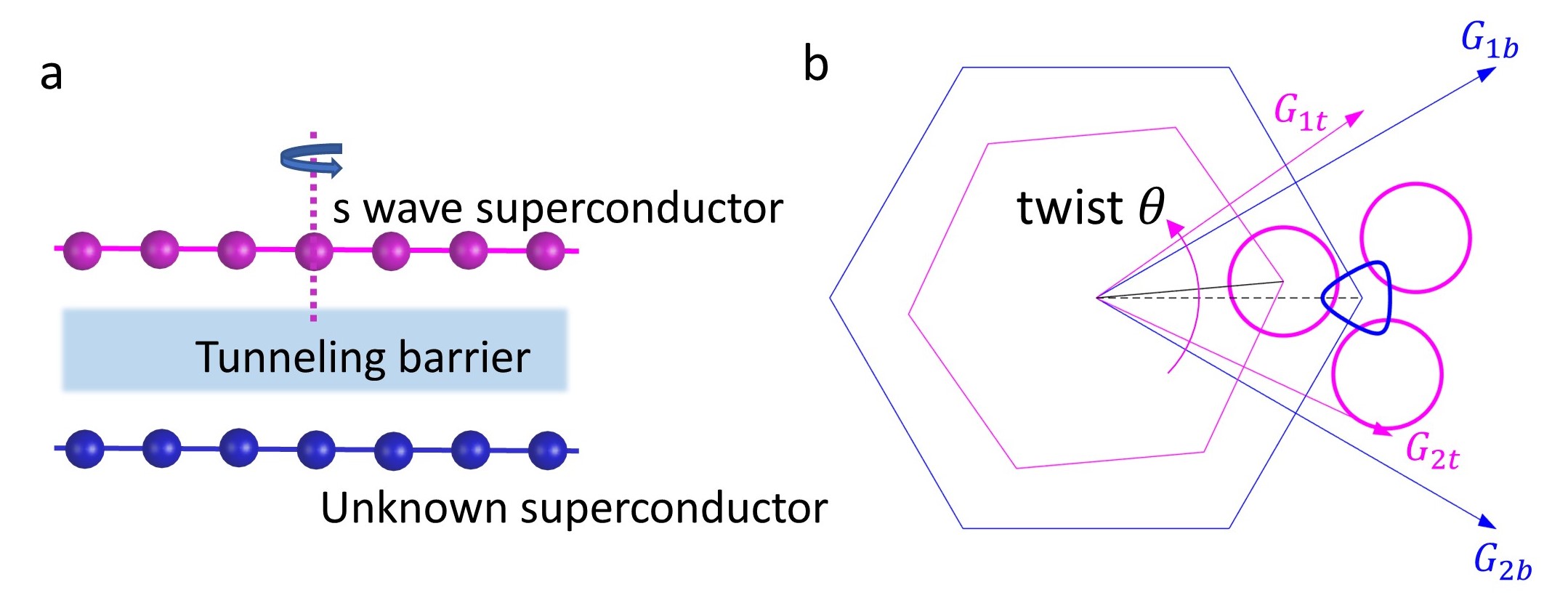}
    \caption{\textbf{Setup for probing the order parameter symmetry by twisted Josephson interferometry.} 
    (a) A Josephson junction between an $s$-wave SC probe (purple) and another SC system (blue) with an unknown order parameter symmetry. The two SCs are separated by a tunneling barrier. Both are assumed to have $C_3$ rotation symmetry. 
    (b) Fermi surfaces (FSs) and Brillouin zones alignment of the probe (top layer, purple) and system (bottom layer, blue). Reciprocal lattice vectors: $\mathbf{G}_{1,t/b}$ and $\mathbf{G}_{2,t/b}$, are denoted by purple/blue arrows respectively. The Bragg scattered FSs of the probe (purple) are plotted around the system FS (blue) $C_3$ symmetrically. When twisting the probe, the probe FS is rotating and intersecting the system FS at different momenta. }
    \label{figure1}
\end{figure}

Here, we propose a general setup for probing the pairing order parameter symmetry of 2D SCs by twisted Josephson interferometry, utilizing various symmetries of the system. As shown schematically in Fig. \ref{figure1}a, the system is composed of an $s$-wave SC as a probe and another SC with an unknown order parameter symmetry in the other side of the junction.
We focus on materials with $C_3$ symmetry, such as TMDs and graphene-based systems.
Assuming momentum resolved tunneling between the two layers, we demonstrate signatures of different order parameter symmetries, such as the twist angle dependence of the critical current, the current-phase relations, and the magnetic field dependence of the Josephson coupling.  
% For instance, nodal order parameters can exhibit a sharp drop of the first order Josephson coupling versus the twist angle.
For example, combining a chiral order parameter SC with an $s$-wave probe generates a dominant third harmonic in the current-phase relation. However, applying a small in-plane Zeeman field breaks the $C_3$ symmetry and creates a linear-in-field first-harmonic Josephson coupling.
As a concrete setup, we study 
%We demonstrate these principles in 
a twisted junction between NbSe$_2$ as an $s$-wave probe and magic angle twisted bilayer graphene (MATBG) as the SC with an unknown order parameter symmetry.  
%into a twisted Josephson interferometer, where the MATBG serves as an unknown SC that we intend to probe its orbital properties. 
% Depending on the order parameter of MATBG, different signatures are revealed and can be potentially detected by experiments, for example, fractional Shapiro steps and Josephson diode effect. 
%Beyond the orbital symmetry, the momentum and spin resolved Josephson couplings could further differentiate the pairing channel between different flavors. \textcolor{blue}{(The last sentence is unclear to me)}

The rest of this article is organized as follows:
In Sec. \ref{II}, we present a symmetry argument based on the Ginzburg-Landau (GL) theory and a microscopic weak coupling model. In Sec. \ref{III}, we demonstrate the probing principle in the twisted NbSe$_2$ and MATBG junction. Sec. \ref{IV} presents in-plane magnetic field dependent Josephson couplings, and Sec. \ref{V} proposes several experimental probing methods. 

% \begin{tikzpicture}
% \tikzset{level distance=2cm}
% \tikzset{sibling distance=1cm}
% \tikzset{every tree node/.style={anchor=base west}}
% \Tree[.\node (a) {measure at $0^{\circ}$ twist};
%          [.\node (b) {};
%              [.\node (d) {d}; ]
%              [.\node (e) {e}; ] ]
%          [.\node (c) {c};
%              [.\node (f) {f}; ]
%              [.\node (g) {g}; ] ] ]
% \end{tikzpicture}
% \begin{figure}
%     \centering
%     \includegraphics[width=\linewidth]{Fig1.jpg}
%     \caption{\textbf{A General principle of probing the order parameter symmetry by twisted Josephson interferometry} 
%     (a) Schematics of the Josephson junction between an $s$-wave SC probe (purple) and another SC system (blue) with unknown order parameter symmetry. Insulating layers are inserted in between to prevent hybridizations between two different materials.
%     (b) Fermi surfaces (FSs) and Brillouin zones alignment of the probe (top layer, purple) and system (bottom layer, blue). Reciprocal lattice vectors: $\mathbf{G}_{1,t/b}$ and $\mathbf{G}_{2,t/b}$, are denoted by purple/blue arrows respectively.  Bragg scattering is considered such that Bragg scattered FSs of the probe (purple) are plotted around the system FS (blue) $C_3$ symmetrically. When twisting the probe, the probe FS is rotating and intersecting the system FS at different momenta. }
%     \label{figure1}
% \end{figure}
\section{Model and symmetry arguments}\label{II}
\subsection{Ginzburg-Landau theory}
Here, we focus on the twisted Josephson junction in which both SCs are invariant under $C_3$ rotation symmetry and time-reversal symmetry (TRS). These symmetries apply to our primary example of a junction between graphene and TMD-based SCs. The superconducting order parameter belongs to one of the following irreducible representations of the $C_3$ group: $A$ (one-dimensional representation, $s$-wave like) or $E$ (two-dimensional representation, $(p_x,p_y)$ or $(d_{x^2-y^2},d_{xy})$ like).\footnote{In cases where there is an additional $C_2$, the $A$ representation breaks into $A_1$ (even under $C_2$, s-wave like) and $A_2$ (odd under $C_2$, f-wave like), and similarly, $E$ breaks into $E_1$ (even, $(d_{x^2-y^2},d_{xy})$ like) and $E_2$ (odd, $(p_x,p_y)$ like). These states remain distinct even in the absence of $C_2$ symmetry if the system is spin rotationally invariant, since one of them is a spin singlet and the other is a triplet.} We assume that spin-orbit coupling is present (provided by the TMD) and that there is no inversion symmetry; hence, singlet and triplet superconductivity are not distinct. 
For simplicity, we will refer to the two representations as $s-$wave or $(p_x,p_y)-$wave, respectively. In the $(p_x,p_y)$ case, we distinguish chiral ($p_x\pm ip_y$) and nematic ($\alpha p_x+ \beta p_y$, with $\alpha,\beta\in \mathbb{R}$) states. If the two SCs forming the junctions have mirror symmetry with respect to a common vertical plane, further distinctions are possible. The different order parameters considered in this work are summarized in Table \ref{table1}.

% and $\Delta_s$, $\Delta_{p\pm ip}$ %(Table $I$). 
% We also include the nodal order parameter $\Delta_{p_x}$ and $\Delta_{p_y}$ for the $C_3$ breaking nematic order case. 
    % \begin{table}[h]
    %     \centering
    %     \begin{tabular}{c|c|c|c}
    %     \hline
    %      Group & IR & transformation under $C_3$ & SC order parameter \\
    %      \hline
    %      \multirow{2}{*}{$C_3$} & A & 1 & $\Delta_s$\\
    %      & E & $e^{\pm i\frac{2\pi}{3}}$ & $\Delta_p^\pm =\Delta_p^x\pm i\Delta_p^y$\\
    %      \hline
    %      \multirow{2}{*}{$D_1$} & A & 1 & $\Delta_s$, $\Delta_p^x$\\
    %      & B & -1 & $\Delta_p^y$\\
    %      \hline
    %     \end{tabular}
    %     \caption{Irreducible representations of the $C_3$ and $D_1$ groups}
    %     \label{tab:my_label}
    % \end{table}
    
The lowest-order symmetry allowed Josephson coupling terms between SCs with either of these order parameters and an $s$-wave SC are:
\textbf{a}. In the $s-$wave to $s-$wave case, both order parameters form the trivial representation under $C_3$ and the first order term $\Delta_{s}^*\Delta_{s}$ is allowed.
\textbf{b}. $s-$wave to $p\pm ip-$wave case: the $C_3$ symmetry is compatible with the chiral order parameter. Under $C_3$ operations, the $n^{th}$ order term $(\Delta_{s}^*\Delta_{p_\pm})^n$ accumulates a phase factor $e^{\pm i\frac{2\pi}{3}n}$. Therefore the lowest order term coupling is $(\Delta_{s}^*\Delta_{p_\pm})^3$, regardless of the rotation angle between two layers. Given the fact that third order coupling is parametrically smaller than the first order coupling in the perturbative tunneling regime, if we break the $C_3$ rotation symmetry it is possible to induce a first order coupling larger in magnitude than the existing third order.
% Given the third order effect is small, if we slightly break the $C_3$ rotation symmetry, it can generate the first order component, which is parametrically stronger than the third order effect and relatively easier to detect in experiment.
Consider breaking of the $C_3$ symmetry by an externally applied magnetic field $\boldsymbol{B}$,
% Here we consider applying the in-plane magnetic field to break the $C_3$ symmetry. 
% Regardless of the microscopic symmetry breaking process (detailed in Section IV), 
we can write down the coupling terms in the Ginzburg-Landau (GL) free energy:
\begin{equation}\label{Eq1}
\begin{aligned}
    F &= \alpha_B \Delta_s^{*} (\boldsymbol{\Delta_p} \cdot \boldsymbol{B}) + \beta_B \Delta_s^{*} (\boldsymbol{\Delta_p} \times \boldsymbol{B})\cdot\boldsymbol{\hat{z}} \\
    &+ \gamma (\Delta_s^* \Delta_p^+)^3 + \gamma (\Delta_s^* \Delta_p^-)^3  + c.c.,
\end{aligned}
\end{equation}
where $\gamma$ denotes the third order $C_3$ symmetric coupling coefficient. 
$\alpha_B$ and $\beta_B$ are the real coupling coefficients (Section IV). 
%Thus, we can expect the magnetic field dependent first order Josephson couplings. 
\textbf{c}. $s-$wave to $p_{x,y}-$wave case: the nodal order parameter breaks $C_3$ spontaneously, therefore $C_3$ symmetry should not be respected by the GL theory in this phase. The first order term $\Delta_{s}^*\Delta_{p_{x,y}}$ is allowed for general twist angles.
If we consider a case where both materials have mirror symmetry, the Josephson coupling should respect the mirror symmetry ($D_1$) when the two mirror planes are aligned. 
Under this condition, for the mirror-symmetric order parameter (denoted by $\Delta_{p_x}$, mirror plane $xz$), the lowest order coupling is still $\Delta_{s}^*\Delta_{p_x}$.
For a mirror anti-symmetric order parameter (denoted by $\Delta_{p_y}$), the first order coupling is forbidden, as it accumulates a $\pi$ phase under the mirror. The lowest order of coupling is $(\Delta_{s}^*\Delta_{p_y})^2$.
Similarly to the chiral case, we can induce a first order harmonic by coupling to an external field.
The symmetry constrained Josephson couplings are summarized in table \ref{table1}.

   \begin{table}
        \caption{ Different order parameters and the corresponding lowest order Josephson couplings. }
        \label{table1}
        \begin{tabular}{c|c|c}
        \hline
        probe & system & Josephson coupling \\
        \hline
        $s$ & $s$ & $\Delta_{s}^*\Delta_{s}$  \\
         \hline
         $s$ &  $p\pm ip$ & $(\Delta_{s}^*\Delta_{p_\pm})^3$ \\
         \hline
         $s$ &  $p_{x,y}$, no mirror & $\Delta_{s}^*\Delta_{p_{x,y}}$ \\
         \hline
         $s$ &  $p_x$, mirror & $\Delta_{s}^*\Delta_{p_{x}}$ \\
         \hline
          $s$ & $p_y$, mirror & $(\Delta_{s}^*\Delta_{p_{y}})^2$ \\
          \hline
        \end{tabular}
    \end{table}

\subsection{Microscopic weak coupling model}
The above symmetry argument is general for different types of SCs. 
Here we consider a simplified microscopic weak coupling model to quantitatively describe Josephson couplings. 
This toy model well captures generic features of the twisted interface between TMDs (e.g., NbSe$_2$) and graphene-based SCs. 
A concrete example: NbSe$_2$ and MATBG twisted heterostructure, is considered in the next section. It shows consistent features compared to the toy model but gives quantitative magnitudes.
Assuming momentum and spin conserving single-electron tunneling element and singlet pairing, the toy model Hamiltonian is:
\begin{align}
    &\mathcal{H}=\mathcal{H}_t+\mathcal{H}_b+{T}, \\
    &\mathcal{H}_l=\frac{1}{2}\sum_\mathbf{p}\Psi^\dag_{l,\mathbf{p}} \begin{pmatrix}
    \epsilon_l(\mathbf{p},\sigma)-\mu_l & i\sigma_y\Delta_l(\mathbf{p})\\
    -i\sigma_y\Delta_l^*(\mathbf{p}) & -(\epsilon_l^T(\mathbf{p},\sigma)-\mu_l) 
    \end{pmatrix}\Psi_{l,\mathbf{p}}, \\ 
    &{T}=\sum_{\{\mathbf{G}_t,\mathbf{G}_b\}} \sum_{\mathbf{p},\sigma}t_{(\mathbf{p}+\mathbf{G}_t)}
    c_{t,\mathbf{p},\sigma}^\dag c_{b,\mathbf{p}',\sigma}
    +h.c.,
\end{align}
where $\Psi^\dag_{l,\mathbf{p}}=(c_{l,\mathbf{p},\sigma}^\dag, c_{l,\mathbf{p},\Bar{\sigma}}^\dag, c_{l,\mathbf{-p},\sigma}, c_{l,\mathbf{-p},\Bar{\sigma}})$, and $c_{l,\mathbf{p},\sigma}$ annihilates a state with spin $\sigma$ and momentum $\mathbf{p}$ in layer $l$ ($l=t,b$). The momentum $\mathbf{p}'$ is determined by: $\mathbf{p} + \mathbf{G}_t = R_\theta(\mathbf{p}' + \mathbf{G}_b$). $\{\mathbf{G}_t,\mathbf{G}_b\}$ are the reciprocal lattice vectors in both layers (marked in Fig. 1), and $R_\theta$ is a c-axis rotation matrix ($\theta$ is the twist angle between two materials).
In our convention, the top layer $l=t$ is the probe (with an $s-$wave order parameter), and the bottom layer $l=b$ is the measured layer (with an unknown order parameter). 
%The possibility of triplet pairing is addressed in section VI. 
The tunneling element $t_{|\mathbf{p}|}$ is assumed to decay fast with the in-plane momentum \cite{bistritzer2011moire}. 

% Probe layer $1$ is twisted with respect to layer $2$ by angle $\theta$, as illustrated in Fig. 1a. The twisted junction also includes an insulating layer in between, which suppress the possible interlayer hybridization between two different materials. The details of the insulating layer does not enter the model, and we assume it tunes the interlayer tunneling amplitude $t$. 
% We consider the interlayer tunneling process $T$ as spin and momentum conserving, i.e., coherent. $\mathbf{p}$ is the momentum in the top probe layer, which couples to the momentum $\mathbf{p}'$ in the bottom layer by Bragg scatterings: $\mathbf{p} + \mathbf{G}_1 = \mathbf{p}' + \mathbf{G}_2$. $\{\mathbf{G}_1,\mathbf{G}_2\}$ is the set of reciprocal lattice vectors in the top and bottom layer. Since the interlayer tunneling amplitude $t_{\mathbf{p+G_1}=\mathbf{p}'+\mathbf{G_2}}$ decays exponentially with the in-plane momentum, we only consider electron tunneling within the range of the first Brillouin zone of the bottom layer $2$. We also assign the real lattice constants of the probe layer to be NbSe2 (purple) and the system layer (blue) to be graphene, which gives the actual Bragg scatterings. 

\begin{figure*}
    \centering
    \includegraphics[width=\linewidth]{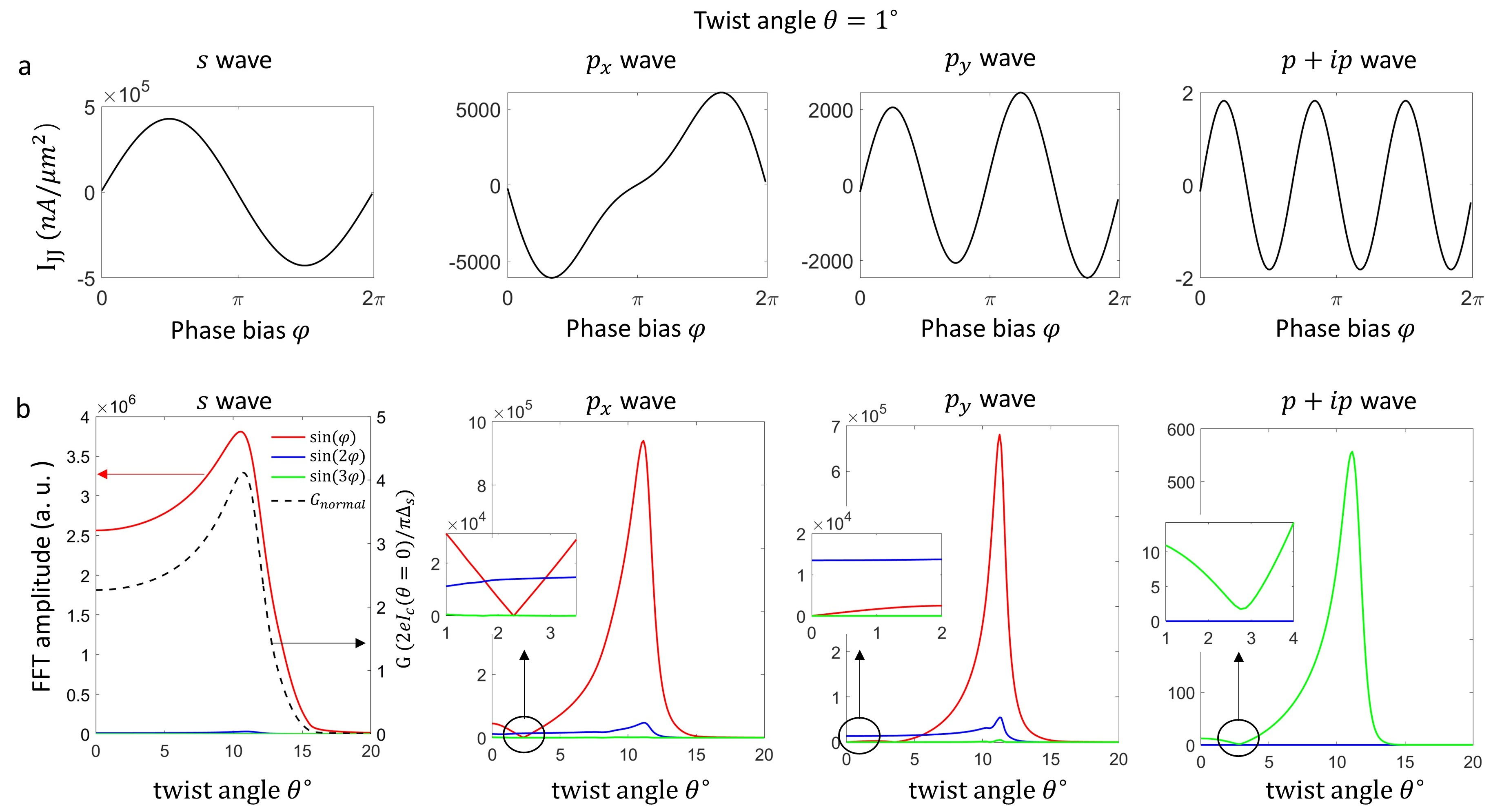}
    \caption{\textbf{Toy model results.} 
    (a) Josephson current at twist angle $\theta = 1^\circ$ for $s$, $p_x$, $p_y$ and $p+ip$ order parameters in the bottom layer. We used the following values for the model parameters: $\lambda_t=5$ eV$\cdot \mathring{A}^2$, $\lambda_{b0}=1$ eV$\cdot \mathring{A}^2$, $\lambda_{b1}=1.5$ eV$\cdot \mathring{A}^3$. The interlayer tunneling amplitude is $t=1$ meV. Chemical potentials are set to be $\mu_t = 600$ meV and $\mu_b=40$ meV. The gaps are $|\Delta_s| = |\Delta_k| = 5$ meV.
    (b) Fast Fourier transform (FFT) amplitude of the current phase relation versus the twist angle $\theta$. FFT amplitudes of different harmonics up to the third order are plotted for $s$, $p_x$, $p_y$, and $p+ip$ cases. In the $s$-wave panel, we plot the normal state conductance $G$ as a function of the twist angle. $G$ is plotted in units of $\frac{2eI_c(\theta=0^\circ)}{\pi\Delta_s}$. Inset of the $p_x$-wave case: the first order harmonics vanish around twist angle $2.3^\circ$ and the second order term becomes the leading order. Inset of the $p_y$-wave case: the first order harmonics vanish at zero twist angle and then start rising when rotating away. Inset of the $p+ip$ wave case: there is a local minimum of critical current in the third harmonics, but it does not vanish. }
    \label{figure2}
\end{figure*}
For the numerical analysis, we assume the following low-energy dispersion: $\epsilon_t(\mathbf{p}=\mathbf{K}_t+\mathbf{k},\sigma) = \lambda_t |\mathbf{k}|^2$ for the probe layer, and $\epsilon_b(\mathbf{p}=\mathbf{K}_b+\mathbf{k},\sigma) = \lambda_{b0} |\mathbf{k}|^2 + \lambda_{b1} k_x (k_x^2-3k_y^2)$ for the measured layer.  The momentum $\mathbf{k}$ here is relative to the $\mathbf{K}_l$ point of each layer $l$. The spectrum near the other valley $\mathbf{K}_l'=-\mathbf{K}_l$ is directly related by TRS.
We use the lattice constants of NbSe$_2$ for the probe layer and graphene for the measured layer to determine the relative position and twist of valleys.
Due to the fast decay of $t_{|\mathbf{p}|}$, we consider only Bragg scattering events within the first Brillouin zone (BZ) of the measured layer.
The Fermi surfaces (FSs) of the probe (purple) and measured (blue) layers are shown in Fig. 1b, with a twist angle of $\theta=5^\circ$.
This picture is in the normal state, without interlayer tunneling ($\Delta=0, t=0$). Turning on the pairing potential opens a gap at the FS. 
With the momentum-resolved tunneling, these band crossing points near the Fermi-level contribute strongly to the phase dependence of the free energy (see Eq. \ref{Eq5}). 
% This "momentum resolved" aspect of Josephson coupling contains the momentum information of order parameters, which can be potentially a probe of the order parameter symmetry.

With the toy model Hamiltonian, we can write down the first and higher order Josephson couplings explicitly, for different order parameter symmetries.
% \begin{equation}
% \begin{aligned}
%      G^{-1} &= G_0^{-1} + T \\
%             &= 
%     \left(
%     \begin{array}{cccc}
%      G_2^{-1} (\mathbf{k}) & T_1(\mathbf{k},\mathbf{k}_1) & T_2(\mathbf{k},\mathbf{k}_2) & T_3(\mathbf{k},\mathbf{k}_3)  \\
%     T_1'(\mathbf{k},\mathbf{k}_1) & G_1^{-1} (\mathbf{k_1}) & 0 & 0 \\
%     T_2'(\mathbf{k},\mathbf{k}_2) & 0 & G_1^{-1} (\mathbf{k_2}) & 0  \\
%     T_3'(\mathbf{k},\mathbf{k}_3) & 0 & 0 & G_1^{-1} (\mathbf{k_3}) \\
%     \end{array}
%     \right) 
% \end{aligned}
% \end{equation}
%  $T$ is the off-diagonal part is the interlayer tunneling matrix, and diagonal part is 
% $G_0^{-1}$ with $G_i(\mathbf{k}) = H_i(\mathbf{k})^{-1}$. Each state $G_2^{-1} (\mathbf{k})$ in the system layer is coupled to three states $G_1^{-1} (\mathbf{k_{1,2,3}})$ in the probe. 
Expanding the free energy to the second order in the tunneling element $t$ (Appendix A), the leading first harmonic component is:
% \begin{align}
% F = k_BT \mathrm{Tr}[\log[G_0^{-1}]] - k_B T \sum_{n} \frac{1}{2n} \mathrm{Tr}[G_0 T]^{2n}
% \end{align}
% The second order term is
% \begin{widetext}
\begin{equation}\label{Eq5}
\begin{aligned}
F^{(1)} &=  \frac{1}{\beta} \sum_{i,\mathbf{k}} \mathrm{Tr}[G_b(\mathbf{k}) T_i G_t (\mathbf{k}_i) T^{\dagger}_i]  \\
&= -\sum_{i,\mathbf{k},n} \frac{2|t_{\mathbf{k}}|^2 |\Delta_{s}| |\Delta_{\mathbf{k}}| \cos{(\varphi+\alpha_{\mathbf{k}})}}{\beta(|\Delta_{\mathbf{k}}|^2+\omega_{n}^2+\xi^2_{b,\mathbf{k}})(|\Delta_{s}|^2+\omega_{n}^2+\xi^2_{t,\mathbf{k}_{i}})} \\
&\stackrel{(\beta=\infty)}{=} -\sum_{i,\mathbf{k}} \frac{|t_{\mathbf{k}}|^2|\Delta_{s}| |\Delta_{\mathbf{k}}| \cos{(\varphi+\alpha_{\mathbf{k}})}}{E_{b,\mathbf{k}}E_{t,\mathbf{k}_i} (E_{b,\mathbf{k}} + E_{t,\mathbf{k}_i})},
        %&= \sum_{k,i} \frac{t^2}{16E_0E_i}(\frac{\tanh{\frac{\beta E_i}{2}}-\tanh{\frac{\beta E_0}{2}}}{E_0-E_i}+\frac{\tanh{\frac{\beta E_i}{2}}+\tanh{\frac{\beta E_0}{2}}}{E_0+E_i})|\Delta_s| |\Delta_{\mathbf{k}}| \cos{(\varphi+\alpha_{\mathbf{k}})}
\end{aligned}
\end{equation}
% \end{widetext}
% \textcolor{blue}{Maybe only show this eq7 and move the details to appendix}
where $\omega_n = (2n+1)\pi /\beta$ is the Matsubara frequency and $\beta=1/k_BT_\text{temp}$ ($T_\text{temp}$ is the temperature). $G_l (\mathbf{k})$ is Green's function on layer $l$. $\xi_{l,\mathbf{k}} = \epsilon_{\mathbf{k}_l}-\mu_l$ and $E_{l,\mathbf{k}} = \sqrt{\xi_{l,\mathbf{k}}^2 + |\Delta_{l,\mathbf{k}}|^2}$. $\Delta_s$ is the top probe layer order parameter (assume to be momentum independent) and $\Delta_{\mathbf{k}}$ is the bottom layer order parameter. $\varphi$ is the relative phase between top and bottom SCs. $\alpha_{\mathbf{k}}$ is the momentum-dependent phase of the unknown order parameter, for instance, $\alpha_\mathbf{k}^{(s)}=0$ for $s$ wave, $\alpha_\mathbf{k}^{(\text{nodal } p)} = \arg(\rm{sgn}(\mathbf{k} \cdot \mathbf{n}))$ (where $\bm{n}$ is the vector along the nodal direction) and $\alpha_\mathbf{k}^{(p\pm ip)}=\arg(k_x\pm ik_y)$ for $p\pm ip$ wave. $i\in{1,2,3}$ is the Bragg scattering summation, where three processes are relevant within the first BZ of the measured layer. 
 %These processes relate momentum $\mathbf{k}$ in the measured layer to momentum $\mathbf{k}_i$ in the probe layer. 

In Eq. \ref{Eq5} we see that each $\mathbf{k}$ point, gives a finite contribution to the first harmonic. However, a nontrivial $\mathbf{k}$ dependent phase $\alpha_{\mathbf{k}}$, creates an interference effect upon integration over momenta and can lead to a vanishing coupling. For example, in the $p+ip$ case, using $C_3$ symmetry and the identity $\sum_{n=0}^2 \cos{(\varphi+\frac{2\pi}{3}n)}=0$, Eq. \ref{Eq5} gives a vanishing first order coupling, as expected from table \ref{table1}. 
% At zero temperature limit, we have (appendix A):
% % \begin{widetext}
% \begin{equation}
% \begin{aligned}
% F^{(2)} 
% \approx \sum_{k,i}& \frac{t^2|\Delta_s| |\Delta_{\mathbf{k}}| \cos{(\varphi+\alpha_{\mathbf{k}})}}{16E_0E_i (E_0 + E_i)}  \\
% \approx \sum_{\mathbf{k}_{c},i} &
% \frac{t^2 \cos{(\varphi+\alpha_{\mathbf{k}_{c}})}}{16(|\Delta_s|+\Delta_{\mathbf{k}_{c}}|)} \Big[ 1 - 
% \frac{2|\Delta_s|+|\Delta_{\mathbf{k}_c}|}{2|\Delta_s|+2|\Delta_{\mathbf{k}_c}|}\frac{(\bm{v_{f1}} \cdot \bm{\delta k})^2}{|\Delta_s|^2} \\ &-
% \frac{|\Delta_s|+2|\Delta_{\mathbf{k}_c}|}{2|\Delta_s|+2|\Delta_{\mathbf{k}_c}|}\frac{(\bm{v_{f2}} \cdot \bm{\delta k})^2}{|\Delta_{\mathbf{k}_c}|^2}\Big]
% \end{aligned}
% \end{equation}
% % \end{widetext}
% $E_0 = \sqrt{\xi_{\mathbf{k}}^2+|\Delta_\mathbf{k}|^2}$ and $E_i = \sqrt{\xi_{\mathbf{k}_i}^2+|\Delta_s|^2}$. If there is a set of band crossing point $\mathbf{k}_c$ in normal states, then we have $\xi_{\mathbf{k}} = 0$ and $\xi_{\mathbf{k}_i}=0$ at the same time. Around these crossing points, it dominantly contributes to the total Josephson current due to the small denominator $E_0 \approx |\Delta_{\mathbf{k}_c}| (1+\frac{(\bm{v_{f2}} \cdot \bm{\delta k})^2}{2|\Delta_{\mathbf{k}_c}|^2})$ and $E_i \approx |\Delta_s| (1+\frac{(\bm{v_{f1}} \cdot \bm{\delta k})^2}{2|\Delta_\mathbf{s}|^2})$, which leads to eq (7). $v_{fl}$ is the Fermi velocity in layer $l$ around crossing points and $\bm{\delta k}$ is the momentum relative to $\mathbf{k}_c$.
% The Josephson coupling thus reflects the phase-sensitive summation of momentum resolved $\Delta_{\mathbf{k}_c}$.
For higher order terms, the nontrivial phase $\alpha_{\mathbf{k}}$ similarly enters the momentum integration and determines the leading harmonics (Appendix A).
% In the next order expansions, we have:
% \begin{widetext}
% \begin{equation}
% \begin{aligned}
% F^{(2)} &= -\frac{k_B T}{4} \sum_{ij} \mathrm{Tr}[G_1(\mathbf{k}) T_i G_2 (\mathbf{k_i}) T_i G_1(\mathbf{k}) T_j G_2 (\mathbf{k_j}) T_j] \\
% &=\sum_{n,k,i,j} \frac{k_BT}{2} \frac{|\Delta_s|^2|\Delta_{\mathbf{k}_i}|^2\cos{(2\varphi + 2\alpha_{\mathbf{k}})}}
% {(|\Delta_\mathbf{k}|^2+\epsilon_n^2+\xi_{\mathbf{k}}^2)(|\Delta_s|^2+\epsilon_n^2+\xi_{\mathbf{k}_i}^2)(|\Delta_\mathbf{k}|^2+\epsilon_n^2+\xi_{\mathbf{k}}^2)(|\Delta_s|^2+\epsilon_n^2+\xi_{\mathbf{k}_j}^2)}
% \end{aligned}
% \end{equation}
% \end{widetext}
% Here we only keep the second order Josephson term, but it also generate the first order Josephson components. $i$ and $j$ are Bragg scattering summations. It describe electrons tunneling from $G_1(\mathbf{k})$ state to  $G_2 (\mathbf{k_i})$ by Bragg scattering $i$ process. Then it tunnels back to $G_1(\mathbf{k})$ but Bragg scattered into $G_2 (\mathbf{k_j})$ and finally goes back to $G_1(\mathbf{k})$.

To quantitatively describe the probing principle, the Josephson current is numerically calculated by:
\begin{equation}\label{Eq6}
\begin{aligned}
    &F = E_0 - \frac{2}{\beta} \sum_{\mathbf{k}, n}\ln{(\cosh{(E_{\mathbf{k} n}/k_BT)})}, \\
    &I_{JJ} = \frac{2e}{\hbar} \frac{dF}{d\varphi},
\end{aligned}
\end{equation}
where $E_0$ is a phase-independent constant and $E_{\mathbf{k} n}$ is the $n^\text{th}$ positive energy eigenvalues of the BdG Hamiltonian $H$ at momentum $\mathbf{k}$. 
Eq. \ref{Eq6} is non-perturbative in the tunneling element $t$, which accounts for the cases when $t\lesssim |\Delta|$.
% The Hamiltonian $H$ includes the inter-layer tunneling element, which means that the numerical results are exact in $t$, this should allow us to account for cases where the inter-layer tunneling amplitude $t$ can be comparable to the order parameter magnitude $|\Delta|$. 
All plotted results are at zero temperature.
% Compared to the perturbative expansion, the calculation includes all orders of the interlayer tunneling processes. Since in reasonable parameter regions, the interlayer tunneling amplitude $t$ can be comparable to the order parameter $\Delta$ strength (appendix).
The Josephson current at twist angle $\theta=1^\circ$ is shown in Fig. \ref{figure2}a, for different order parameter symmetries. 
For the trivial $s-$wave - $s-$wave case, the first harmonic is the leading term, as expected. 
For the nodal order parameter $\Delta_{p_x}$ and $\Delta_{p_y}$, the first order term $\Delta_s^*\Delta_{p_{x/y}}$ is generally allowed but will be suppressed compared to the $s-$wave case, due to the sign changing. 
% Since the order parameter $\Delta_{p_x,p_y}$ changes sign in the momentum space, there will be cancellations of contributions to the first order Josephson current, as seen from Eq.\ref{Eq5}. 
% If the contribution from both lobes of the nodal order parameter is comparable, we would expect the first order coupling to be strongly suppressed, and the second order term from the Copper co-tunneling $(\Delta_s^*\Delta_{p_{x/y}})^2$ to show up in the current-phase relations, since this term add up constructively regardless of the sign change of $\Delta_{p_{x/y}}$. 
If the first order term is strongly suppressed, the second order term from the Copper pair co-tunneling $(\Delta_s^*\Delta_{p_{x/y}})^2$ shows up since this term adds up constructively in momentum space. 
From Fig. \ref{figure2}a, we indeed see a mixture of the first and second harmonics in both $p_x$ and $p_y$ cases. 
Since the toy model has mirror symmetry (mirror plane $xz$) and $\Delta_{p_y}$ is odd under the mirror, it leads to perfect cancellation of the first order component when two mirror planes are aligned ($\theta=0^\circ$). Even twisted $\theta=1^\circ$ away, we still see a strong second order Josephson component.
On the other hand, $p_x$ wave is even under the mirror and the first harmonic is dominant near the zero twist angle. 
For the chiral $p+ip$ order parameter, only $\sin(3\varphi)$ component exists.
% This feature can be used to probe graphene multilayer SCs with mirror symmetry. Near the zero twist angle, the strong second order Josephson coupling can be a signature of a nodal order parameter. 

% The Josephson coupling evolves with the twist angle. 
Fig. \ref{figure2}b shows the twist angle dependence of different harmonic components (first, second and third) in the current-phase relation. 
For different order parameter symmetries, the amplitudes all drop down around a twist angle of $\theta=15^\circ$. This is because two FSs do not overlap for larger twist angles. %rotate further away and there are no available energy-momentum matched states.
We also plot the normal state conductance $G$ versus twist angle, in units of $2eI_c(\theta=0^\circ)/\pi\Delta_s$ (left panel, Fig. \ref{figure2}b). 
From the Ambegaokar–Baratoff (AB) relation \cite{ambegaokar1963tunneling}, the normal state conductance and the critical current between two $s-$wave SCs should obey $I_c/G = \pi \Delta_s/2e$. 
In our case, the in-plane momentum is conserved (contrary to the assumption used in the derivation of the AB relation). Nevertheless, we find that in the $s-$wave case, the ratio $G/I_c$ is approximately constant as the twist angle varies. This is apparent in Fig. \ref{figure2}b, where 
$I_c (\theta)$ and $G(\theta)$ are seen to follow a similar twist angle dependence. This is not the case for non-$s-$wave order parameter symmetries. 
% , which distinguishes from other order parameters.

For the chiral $p+ip$ order parameter, only $\sin(3\varphi)$ component exists for all twist angles, as we predicted by symmetry. The variation of the FFT amplitude depends on the band alignments.
%For the $p_y$ wave, we can see there is no first order component at zero twist angle due to the mirror symmetry (zoom-in panel). But first order starts to grow and later dominates when rotating away. \textcolor{blue}{(this mirror argument is presented also in the previous paragraph. Maybe we should leave only 1 of them)}

\begin{figure*}
    \centering
    \includegraphics[width=\linewidth]{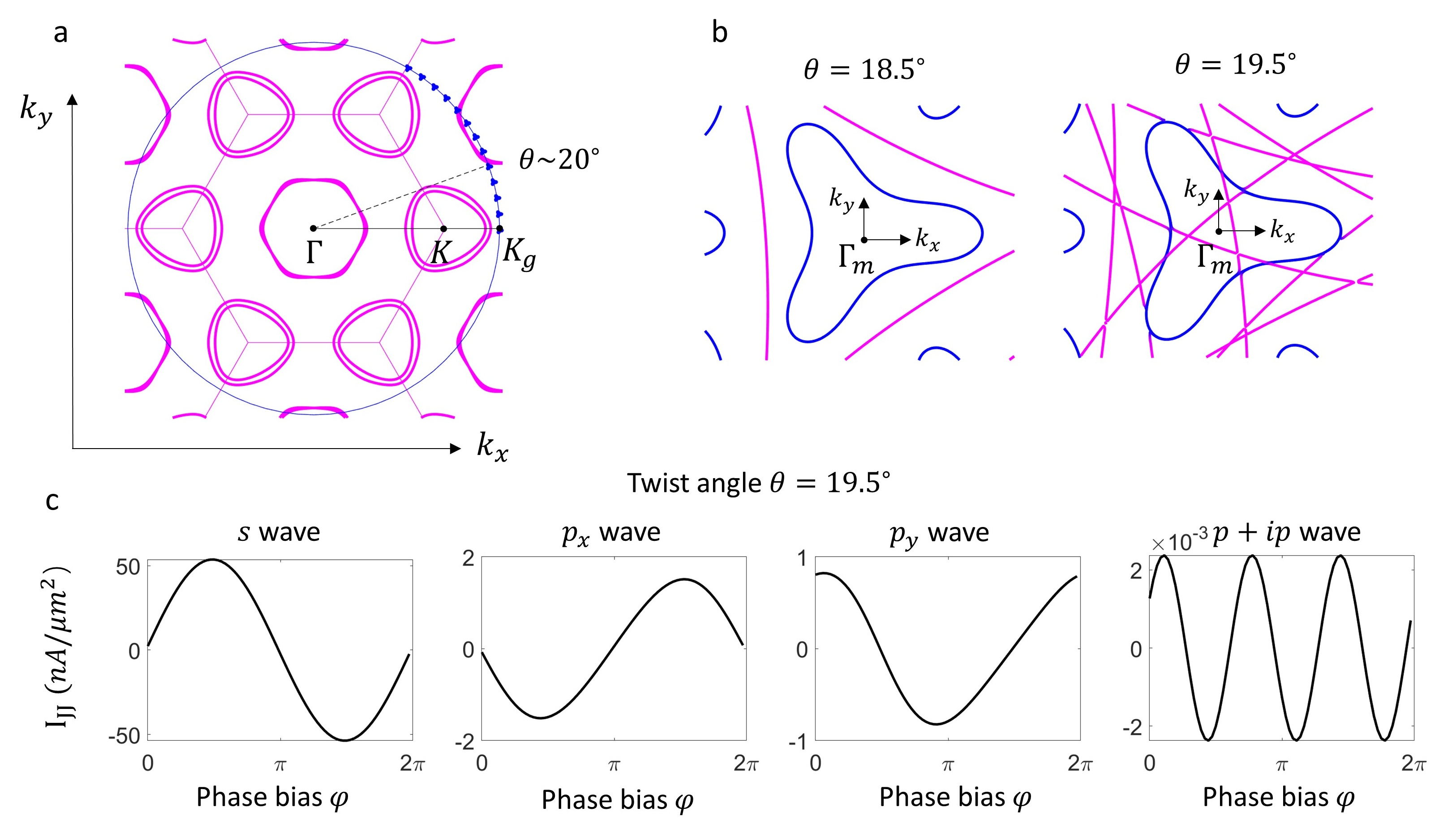}
    \caption{\textbf{NbSe$_2$ - MATBG twisted heterostructure.} 
    (a) FS alignment between NbSe$_2$ and MATBG, over a $60^\circ$ range of rotation angle. Purple pockets are from NbSe$_2$ and small blue pockets are from MATBG. $K$ is the valley of NbSe$_2$ and $K_g$ is the valley of graphene. For MATBG, the Fermi level is set to $\mu=5$ meV relative to charge neutrality, corresponding to a carrier density of $\sim 1.3 \cdot 10^{12} \rm{cm}^{-2}$ (assuming that all four flavors are filled equally).
    (b) Zoom in of the FSs at twist angles $\theta=18.5^\circ$ and $\theta=19.5^\circ$. $\Gamma_m$ is the center of the mini-Brillouin zone.
    (c) The current-phase relation at twist angle $\theta=19.5^\circ$ for different order parameter symmetries in the MATBG: $s$, $p_x$, $p_y$ and $p+ip$.}
    \label{figure3}
\end{figure*}

Noticeably, for the $p_x$ wave case, we see a V-shaped drop of the first order term around twist angle $2.3^\circ$. At this angle, the first order term vanishes due to destructive interference, and the second order coupling is dominant ( Fig. \ref{figure2}b inset). 
The second order term here has a negative sign, which gives a free energy minimum at non-zero phase bias $\varphi\neq 0$, implying TRS breaking \cite{can2021high}. 
Note that this V-shaped drop happens at a generic angle and depends on band alignment details, and not from symmetry considerations. 
It serves as a signature of a sign changing order parameter even in the absence of mirror symmetry.
Close to the perfect vanishing angle, the TRS breaking due to the comparable mixture of first and second harmonics can be detected by the Josephson diode effect \cite{jiang2022superconducting}.
Assuming higher order terms are negligible, there is a one-to-one correspondence between the amount of asymmetry in the critical current to the ratio between the first and second harmonic magnitudes (Appendix G).

\section{Using NbSe$_2$ to probe the order parameter of MATBG}\label{III}
The toy model demonstrates how different order parameter symmetries of the system manifest themselves in the angle-dependent current-phase relation of the twisted junction. In order to provide quantitative predictions, we now study a concrete example: twisted Josephson junction between NbSe$_2$, acting as an $s-$wave SC probe, and MATBG, an SC with an unknown order parameter symmetry. 
The two SCs are separated by two layers of WSe$_2$ that serve as a tunneling barrier.
The barrier suppresses the interlayer hybridization and charge transfer, as confirmed by DFT calculations (Appendix D), such that the MATBG layer is not strongly perturbed by the NbSe$_2$. 
For simplicity, we consider monolayer NbSe$_2$ as a probe. 
%But in principle, few layer and bulk NbSe2, or other $s$ wave superconductors with clean surfaces would also work. 

% \textcolor{blue}{Maybe put the whole paragraph in the appendix.} 
The MATBG layer is described by the continuum model \cite{bistritzer2011moire}.
NbSe$_2$ is considered by a three-orbital tight-binding model with the orbital basis: $d_{z^2,\uparrow/\downarrow}$, $d_{xy,\uparrow/\downarrow}$, and $d_{x^2-y^2,\uparrow/\downarrow}$ \cite{liu2013three}.
The Josephson current is calculated by including a mean field pairing potential in each layer and momentum-resolved tunneling between the two SCs, with a tunneling element $t=0.1$ meV (Appendix B).
% The tunneling includes Bragg scattering processes within the first BZ of the graphene layer (Appendix B), similar to the tunneling in the toy model of the previous section. 

The FSs of the twisted junction are shown in Fig. \ref{figure3}a. NbSe$_2$ has two types of electron pockets (purple) around $\Gamma$ point, $K$, and $K'$ point. The strong Ising spin-orbit coupling (SOC) splits the spin-up and spin-down components. 
The FSs of MATBG (blue) are also plotted, where we show the trajectory of the MATBG FS as the twist angle between the two SCs changes from $0^\circ$ to $60^\circ$. 
% Around $\theta=0^{\circ}$, the NbSe$_2$ $K$ electron pockets are close to the MATBG FS but do not intersect them within the first mini BZ.
Around $\theta=20^{\circ}$, the $\Gamma$ pocket of the NbSe$_2$ from the second BZ intersects the MATBG FS and gives a complicated band alignment, as shown in Fig. \ref{figure3}b for $\theta=18.5^{\circ}$ and $\theta=19.5^{\circ}$.
%It generates much weaker Josephson currents by coupling wave functions in remote mini Brillouin zones to NbSe2.
% Remarkably, when rotating around $\theta=20^\circ$ away, the $\Gamma$ pockets of NbSe$_2$ in the next BZ cross the MATBG bands over a large angle range from $20^\circ$ to $40^\circ$. 
% Since the mini BZ size is small, as controlled by $1.1^\circ$ magic angle, even $1^\circ$ rotation sweeps the NbSe$_2$ electron pockets over the entire MATBG mini BZ and generates completely different band cutting conditions and Josephson couplings. As shown in Fig. 3b, NbSe$_2$ bands starts to approach MATBG at $\theta=18.5^\circ$, then sweeps over and gives a complicated band alignment at $\theta=19.5^\circ$.

\begin{figure*}
    \centering
    \includegraphics[width=\linewidth]{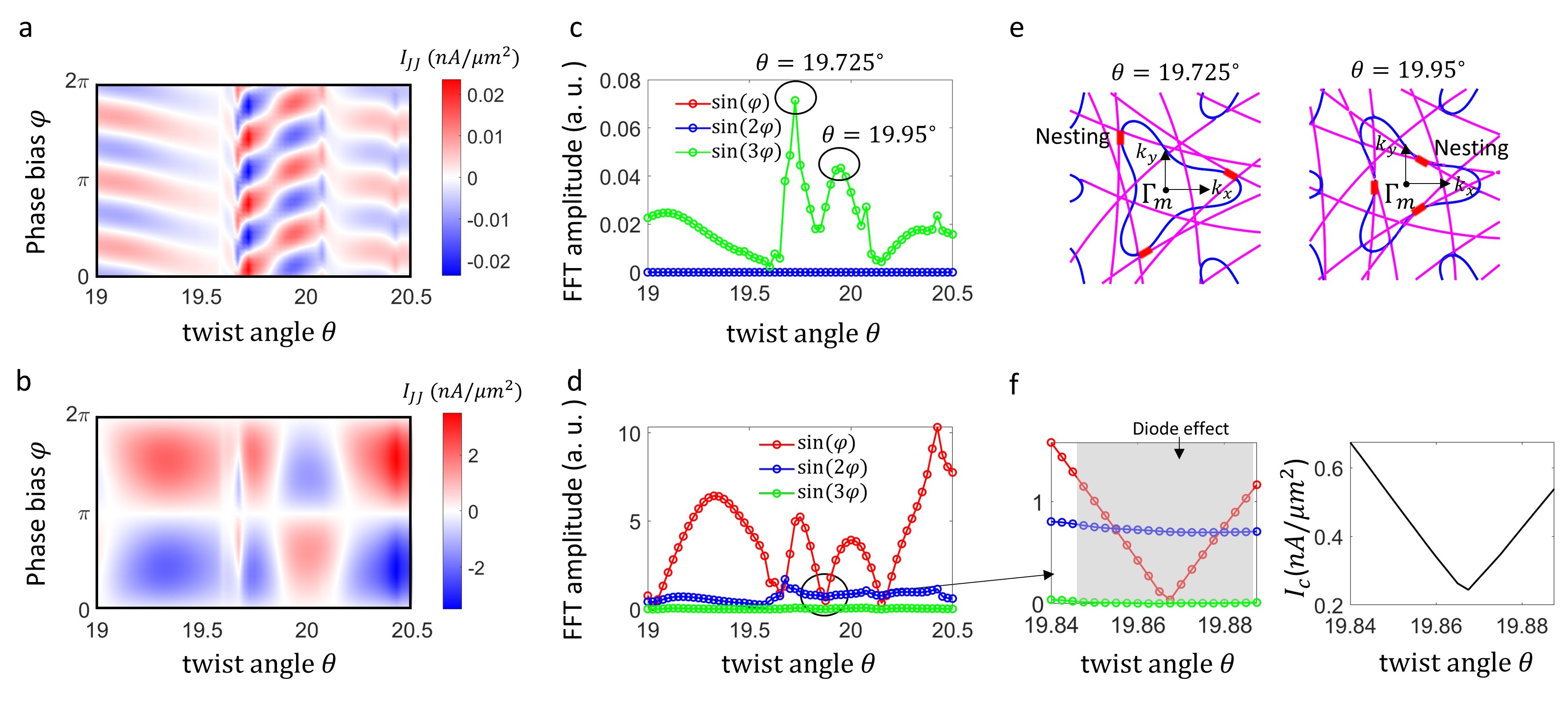}
    \caption{\textbf{Twist angle dependent Josephson current in the NbSe$_2$ - MATBG heterostructure.} 
    (a), (b) Colormap of the Josephson current as a function of twist angle and phase bias for $p+ip$ and $p_x$ order parameter in the MATBG, respectively.
    (c), (d) The FFT amplitude of the current phase relations as a function of twist angle for $p+ip$ and $p_x$ order parameters. 
    (e) The FSs at two twist angles from panel (c). The FSs of NbSe$_2$ and MATBG are tangent to each other (marked in red).
    In (f), a zoom-in FFT amplitude and critical current of the $p_x$ wave order parameter around twist angle $19.88^\circ$ is shown. The first order component vanishes. In the shaded area, the Josephson diode effect can be observed.}
    \label{figure4}
\end{figure*}

Fig. \ref{figure3}c shows the current-phase relations for different order parameter symmetries at twist angle $\theta=19.5^\circ$. An $s-$wave order parameter gives a trivial $\sin(\varphi)$ dependence. The magnitude of the critical current is around 50 $\rm{nA}/\mu m^2$, which is much larger than the critical current for non-$s-$wave pairing symmetries. 
For $p_x$ and $p_y-$wave cases, a mixture of first and second order harmonics is observed, with a magnitude of around 2 $\rm{nA}/\mu m^2$.
%While for the $p_y$ wave, we see the dominant second order harmonics, since twist angle $\theta=20^\circ$ accidentally falls into the dip of V shape drop, as addressed later.
For the chiral case, the lowest order of Josephson coupling is $\sin(3\varphi)$, with a magnitude of around 2 $\rm{pA}/\mu m^2$. 
If the interlayer coupling is increased by using a thinner tunneling barrier or by applying pressure to the junction, the third-order Josephson coupling is strongly enhanced (as it scales as $t^6$). 
For instance, increasing $t$ from $0.1$ meV to $0.4$ meV, Josephson current is increased by more than three orders of magnitude (Appendix C) and can reach the order of few $\rm{nA}/\mu m^2$. 
% Without a tunneling barrier, NbSe$_2$ couples to graphene multilayers strongly, with $t\approx 20$meV \cite{gani2019superconductivity}. This sets an upper limit of the interlayer tunneling strength.
% By varying insulating layer thickness and also applying pressure to the junction, $t$ can be suppressed down to the perturbative region or tuned to large values comparable to the order parameter strength. 
% Even beyond the perturbative region, predictions of higher ha/rmonics generations based on symmetries still apply.

The Josephson coupling has a complicated twist angle dependence. Figs. \ref{figure4}a and \ref{figure4}b show the Josephson current as a function of phase bias and twist angle for the chiral $p+ip$ and nodal $p_x$ wave cases. 
For a $p_x$ order parameter, due to TRS, the current-phase relation is odd with respect to $\varphi \rightarrow -\varphi\,\rm{mod}(2\pi)$, i.e., the free energy $F(\varphi)$ is even. 
This does not apply in the $p+ip$ case, since this order parameter breaks TRS. 
%For the $p_x$ case, TRS can only be broken spontaneously, and therefore the free energy dependence on the relative phase has to be time-reversal symmetric.} 

Figs. \ref{figure4}c and \ref{figure4}d show the corresponding FFT amplitude of different harmonics versus the twist angle from $\theta=19^\circ$ to $\theta=20.5^\circ$. 
For $p+ip$, the lowest order is $\sin(3\varphi)$, but its magnitude varies with the twist angle. In the plotted angle range, we see two strong peaks. These peaks occur when the FSs of MATBG and NbSe$_2$ are tangent to each other.
% These two peaks are also seen in the $p_x$ wave case.
% which actually corresponds to band nesting between MATBG and NbSe$_2$ Fermi surfaces. We can approximate NbSe$_2$ bands as lines. When they sweep over MATBG bands by rotation, it creates nesting conditions (marked as red lines in Fig. 4c) when touching the "lobe" (the first peak) and the "belly" (the second peak) of MATBG Fermi surfaces. 

For the nodal $p_x$ wave case, the first order component generally dominates. There are several sign changes. 
Around them, the FFT amplitude shows sharp V-shaped drops in the first harmonic. 
Fig. \ref{figure4}e shows a zoom-in of one of these drops, where the first order component vanishes but the $\sin(2\varphi)$ component survives. 
%It gives the free energy minimum at the relative phase bias $\phi=\pm\pi/2$, which breaks TRS spontaneously. 
This feature is well captured by the toy model, and occur quite generally here given the complicated band cutting conditions.
As in the toy model, the sign of the second harmonic term is such that it favors time reversal symmetry breaking (similarly to Ref.~\cite{can2021high}). In the shaded region in Fig. \ref{figure4}f, we expect to see a Josephson diode effect \cite{jiang2022superconducting} (Appendix G).
% However, note that, for this specific MATBG case, the system suffers from the angle disorder in real space. If the probe covers several angle disorder regions, the Josephson coupling may be dominated by the $s$ wave part. To avoid this, it is better to use a small SC probe around $100$nm by $100$nm, to only probe a single twist angle domain. 

\section{In-plane magnetic field induced Josephson couplings}\label{IV}
For the chiral order parameter, we have shown a robust $\sin(3\varphi)$ dependence in the current-phase relation, regardless of the twist angle, as long as the $C_3$ symmetry is maintained. 
The third order harmonic scales as the interlayer tunneling amplitude $t$ to the sixth power in the perturbative limit.
On the other hand, if we slightly break the $C_3$ symmetry (either by an in-plane magnetic field or by strain), we can generate a first order component that scales as $t^2$, which can be more significant than the intrinsic third order coupling. 
Phenomenologically, the phase-dependent free energy term is given in Eq. \ref{Eq1}.

Microscopically, an in-plane magnetic field generates both the Zeeman effect and the orbital effect.
For a material with Rashba SOC, the Zeeman effect distorts the energy bands in a way that breaks the $C_3$ symmetry. 
By introducing these effects in the toy model, the Josephson current to the second order in $t$ has the form (Appendix E):
\begin{equation}\label{Eq7}
    \begin{aligned}
       I^{(1)} 
    &= \sum_{\mathbf{k}} [I_{0\mathbf{k}} + I_{1\mathbf{k}}(\mathbf{B})] \cos(\varphi + \alpha_{\mathbf{k}}+\beta_{\mathbf{k},\mathbf{B}}),
    \end{aligned}
\end{equation}
% \begin{widetext}
% \begin{equation}
% \begin{aligned}
%      F^{(2)}_{Zeeman} &= -2k_BT\sum_{n,\mathbf{k},i} \frac{\lambda^2(k_x^2+k_y^2)+\xi_{\mathbf{k}_i}^2+\Delta_{\mathbf{k}_i}^2-B^2}{(\Delta_1^2+\xi_{\mathbf{k}}^2)((\Delta_{\mathbf{k}_i}^2+\xi_{\mathbf{k}_i}^2-\lambda^2(k_x^2+k_y^2-B^2))^2-4\Delta_{\mathbf{k}_i}^2\lambda^2(k_x^2+k_y^2-4\lambda^2(\mathbf{B}\times \mathbf{k})^2))} \Delta_1^*\Delta_{\mathbf{k}_i} \cos(\phi) \\
%      &\approx -(F^{(2)} + \sum_{\mathbf{k}}\alpha_\mathbf{k} \lambda^2(\mathbf{B}\times \mathbf{k})^2 ) \cos(\phi)
% \end{aligned}
% \end{equation}
% \end{widetext}
% The term $\alpha_\mathbf{k}$ transforms like $\Delta_{\mathbf{k}_i}$ under $C_3$, while $(\mathbf{B}\times \mathbf{k})^2$ contains parts the transforms like both chiralities. Therefore, upon summation over $\mathbf{k}$ it will result in a non vanishing contribution  proportional to $B^2$ and $\lambda^2$. Note that it can generate the linear $B$-dependent first order Josephson couplings, which is allowed in symmetry, but occur in the higher order expansions.
where $\alpha_{\mathbf{k}}$ is the order parameter's momentum-dependent phase and $\varphi$ is the phase differences between two SCs. $I_{0\mathbf{k}}$ is independent of the Zeeman field and invariant under $C_3$ rotations.
$I_{1\mathbf{k}}$ is a function of $(\mathbf{k}\cdot \mathbf{B})^2,(\mathbf{k}\times\mathbf{B})^2,|\mathbf{B}|^2$.
There is also a phase shift $\beta_{\mathbf{k},\mathbf{B}} = b_{\mathbf{k}} (\mathbf{k} \cdot \mathbf{B})$, linear in $|\mathbf{B}|$ to the leading order. The coupling coefficient $b_{\mathbf{k}}$ is composed of microscopic parameters, such as, SOC strength and momenta.
For the trivial $s-$wave case, $\alpha_{\mathbf{k}}$ is a constant and the integration over momentum gives a non-vanishing first order at zero field. With an existing first order component, both the phase shift $\beta_{\mathbf{k},\mathbf{B}}$ and $I_{1\mathbf{k}}$ term from the Zeeman field only generates a $|\mathbf{B}|^2$ dependent critical current. 
However, for the chiral order parameter, the first order term vanishes at zero field due to the negative interference in the momentum space. When Zeeman field generates a phase shift $\beta_{\mathbf{k},\mathbf{B}}$, it translates into a linear in $|\mathbf{B}|$ first harmonic component in the current-phase relation (Appendix E). 

\begin{figure}
    \centering
    \includegraphics[width=\linewidth]{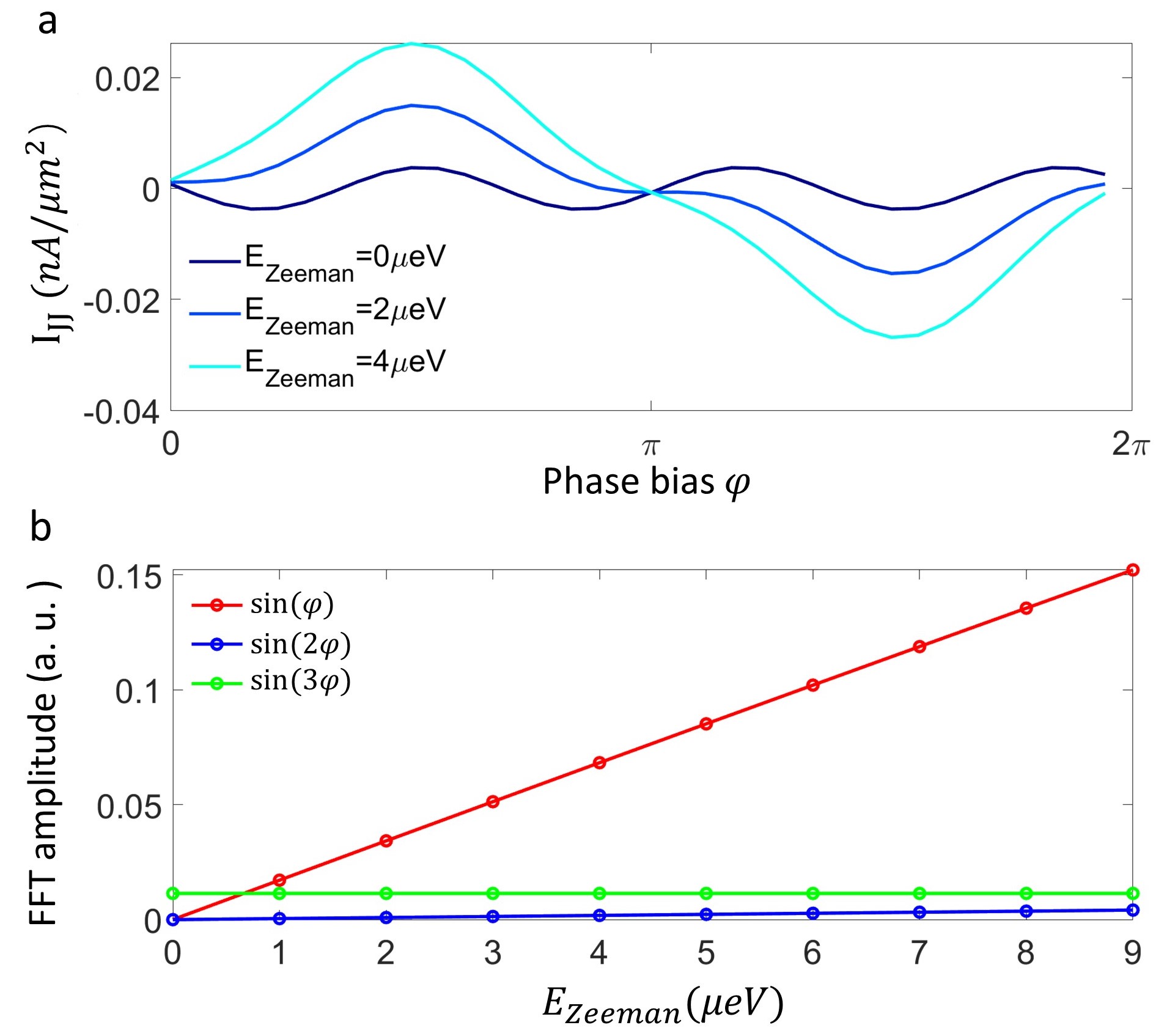}
    \caption{\textbf{Zeeman field induced Josephson coupling for a $p+ip$ order parameter in the probed layer.}
    (a), (b) The current-phase relation at twist angle $\theta=19.5^\circ$ for different Zeeman field strengths.
    (c), (d) The FFT amplitudes of the first, second, and third harmonics of the current-phase relation versus the Zeeman field strength. }
    \label{figure5}
\end{figure}

In the twisted NbSe$_2$-MATBG junction as an example, Figs. \ref{figure5}a and \ref{figure5}b shows the current-phase relation and FFT amplitudes at twist angle $\theta=19.5^\circ$ for different in-plane Zeeman field strengths. A $p+ip$ order parameter is assumed in the MATBG. 
At zero field, there is only a $\sin(3\varphi)$ component. As the field increases, a $\sin(\varphi)$ component is generated, whose amplitude is linear in the field strength. 
% Fig. \ref{figure5}b shows the versus the Zeeman field. 
% First order harmonics grow up with the Zeeman field. We indeed see a linear dependence in $B$, consistent with the previous toy model and Ginzburg Landau free energy.

% To consider the orbital effect from the momentum kick by the gauge-field, one need to be careful with the tunneling element.
% For a fully momentum conserving tunneling, introducing any in-plane magnetic field results in a full decoupling of the SCs. Since, by assuming momentum conservation, we actually assume an infinite junction, and the decoupling of the SCs is a result of the Fraunhofer patterns effect. 
Next, we consider the orbital effect of the magnetic field. In an infinite junction, assuming fully momentum-resolved tunneling, an arbitrarily small in-plane field completely decouples the order parameters of the two SCs. This is because the field creates a momentum mismatch between the two SC order parameters across the junction. 
If we relax the momentum conservation, assuming instead that the tunneling conserves momentum only up to 1/L (where L is the lateral size of the junction), we find a symmetry-breaking-induced first harmonic that is proportional to $\Phi^2$, where $\Phi$ is the flux through the junction. 
This is true for all orders in perturbation theory in $t$ (Appendix E).  Therefore, we expect the linear term related to the interplay of the Zeeman effect and Rashba SOC to dominate at small in-plane fields.

Another way to create an orbital effect that breaks the symmetry is to drive an in-plane supercurrent through one of the SCs. Then, the order parameter in that layer acquires finite momentum $2\mathbf{q}$ which results in a similar effect to that of the in-plane magnetic field. It also generates an additional symmetry-breaking effect from a shift of the energy spectrum of the quasi-particles. However, this effect does not contribute to the first harmonic Josephson coupling in linear order in $\mathbf{q}$ (Appendix E).
% The first-harmonic Josephson coupling is proportional to $t^2\xi^2|\boldsymbol{q}|^2$ where $\xi$ here is the SC coherence length (\textcolor{blue}{Appendix}).

\section{Discussions}\label{V}
We have studied how different order parameter symmetries of two-dimensional SCs manifest themselves in a twisted Josephson junction with an anisotropic $s-$wave SC as a probe. For an $s-$wave order parameter, the critical current dependence on the twist angle is expected to follow closely the dependence of the normal state conductance. Therefore, a strong deviation from this dependence is an indicator of a sign changing order parameter.

The periodicity of the dominant term in the current-phase relation can be directly detected by Shapiro steps in the a.c. Josephson effect. Instead of integer steps in the dc voltage from the current bias: $V_{dc} = n\frac{\hbar \omega_{{ac}}}{2e}$, we expect to see fractional steps: $V_{dc} = \frac{n}{m}\frac{\hbar \omega_{{ac}}}{2e}$ for a $\sin({m\varphi})$ term. A further discussion of the fractional Shapiro steps, including estimates for experimental parameters where they may be observed, is provided in Appendix F.

For nodal order parameters, we predict the suppression of the first order Josephson coupling at generic twist angles (not necessarily dictated by symmetry). The first order coupling shows V-shaped drops versus the twist angle. Around these angles, the energy-phase relation is expected to be dominated by the second harmonic. Generically, the second-order Josephson coupling has a sign that favors TRS breaking (i.e., the minimum of the energy occurs away from $0$ and $\pi$.) \cite{can2021high}. We show that an asymmetry in the critical current is expected in this case (the so-called Josephson diode effect). 
% Moreover, we show that if the energy-phase relation is dominated by the first and second harmonics (with higher order terms being negligible): a) The asymmetry (the ratio between critical current in one direction to the critical current in the opposite direction) is bounded. b) There exist a one to one correspondence between the amount of asymmetry to the ratio between the first and second harmonic magnitudes. (Appendix G) 

The direct measurement of the angle dependent critical current is also interesting. For the specific NbSe$_2$-MATBG case, a strong enhancement of the critical current is predicted between twist angle $\theta=20^\circ$ to $\theta=40^\circ$. In this twist angle range, the large NbSe$_2$ FSs around the $\Gamma$ point in the second BZ intersect the tiny FSs of the MATBG around the K point. A similar situation is expected in other graphene-based superconductors. 
The recently developed quantum twisting microscope \cite{inbar2023quantum} is a promising tool to study the twist angle dependent critical currents and current-phase relations, and identify different order parameter symmetries.
% It is also possible to have an observable Josephson couplings around $\theta=0^\circ$, where the $K$ pockets of NbSe$_2$ are relevant. 
% Even through they are not cutting the first mini BZ of MATBG, they are rather close in momentum and energy to generate Josephson couplings. 

\section{Acknowledgement}
We thank Shahal Ilani for stimulating discussions. This work was supported by the European Research Council (ERC) under the European Union’s Horizon 2020 research and innovation programme (grant agreement No 817799), the US-Israel Binational Science Foundation (BSF), and the CRC 183 of the Deutsche Forschungsgemeinschaft (Project C02).

% \clearpage
\section{Appendix}
\subsection{Higher order expansions from the toy model}
In the toy model from Sec. \ref{II}, the Green's function of the twisted junction is:

\begin{equation}\label{Eq8}
\begin{aligned}
     G^{-1} &= G_0^{-1} + T \\
            &= 
    \left(
    \begin{array}{cccc}
     G_b^{-1} (\mathbf{k}) & T_1(\mathbf{k},\mathbf{k}_1) & T_2(\mathbf{k},\mathbf{k}_2) & T_3(\mathbf{k},\mathbf{k}_3)  \\
    T^\dagger_1(\mathbf{k},\mathbf{k}_1) & G_t^{-1} (\mathbf{k_1}) & 0 & 0 \\
    T^\dagger_2(\mathbf{k},\mathbf{k}_2) & 0 & G_t^{-1} (\mathbf{k_2}) & 0  \\
    T^\dagger_3(\mathbf{k},\mathbf{k}_3) & 0 & 0 & G_t^{-1} (\mathbf{k_3}) \\
    \end{array}
    \right).
\end{aligned}
\end{equation}
The diagonal part $G_l(\mathbf{k}) = H_l(\mathbf{k})^{-1}$, where $H_l$ is the Hamiltonian in layer $l$. $T$ is the off-diagonal part, the interlayer tunneling matrix. We assume $T_i = t\tau_3$, where $\tau_i$ are Pauli matrices that act in Nambu space. 
In the perturbative limit in $t$, we can expand the free energy as:
\begin{align}
F = -\frac{1}{\beta} \mathrm{Tr}[\log[G_0^{-1}]] +\frac{1}{\beta} \sum_{n} \frac{1}{2n} \mathrm{Tr}[G_0 T]^{2n}.
\end{align}
The leading first harmonic term is:
\begin{widetext}
\begin{equation}\label{Eq10}
\begin{aligned}
F^{(1)} &=  \frac{1}{2\beta} \mathrm{Tr}[G_0 T]^{2} =  \frac{1}{\beta} \sum_{n,\mathbf{k},j} \rm{Tr}[G_b (\mathbf{k},i\omega_n)T_jG_t(\mathbf{k}_j,i\omega_n)T^\dagger_j] = F_0^{(1)} - \sum_{n,\mathbf{k},j} \frac{2|t|^2|\Delta_{\mathbf{k}}||\Delta_{s}| \cos(\alpha_{\mathbf{k}}+\varphi)} {\beta(\omega_n^2+|\Delta_{\mathbf{k}}|^2+\xi_{b,\mathbf{k}}^2) (\omega_n^2+|\Delta_s|^2+\xi_{t,\mathbf{k}_{j}}^2) } \\
  &= F_0^{(1)} - \sum_{n,\mathbf{k},j} \frac{|t|^2|\Delta_{\mathbf{k}}||\Delta_{s}|}{E^2_{b,\mathbf{k}}-E^2_{t,\mathbf{k}_j}} \Big[ \frac{f(E_{b,\mathbf{k}})-f(-E_{b,\mathbf{k}})}{E_{b,\mathbf{k}}} - \frac{f(E_{t,\mathbf{k}_j})-f(-E_{t,\mathbf{k}_j})}{E_{t,\mathbf{k}_j}}  \Big]\cos{(\alpha_{\mathbf{k}} + \varphi)} \\
  &\stackrel{(\beta=\infty)}{=}  F_0^{(1)} - \sum_{\mathbf{k},j} \frac{|t|^2 |\Delta_{\mathbf{k}}||\Delta_s|}{ E_{b,\mathbf{k}} E_{t,\mathbf{k}_j} (E_{b,\mathbf{k}}+E_{t,\mathbf{k}_j})} \cos{(\alpha_{\mathbf{k}}+\varphi)},
\end{aligned}
\end{equation}
\end{widetext}
where $\omega_n$ is the Matsubara frequency and $f(E)$ is the Fermi-Dirac distribution function. 
$E_{l,\mathbf{k}} = \sqrt{\xi^2_{\mathbf{k}}+|\Delta_{l}|^2}$, with $l=t/b$. The order parameter phase $\alpha_\mathbf{k}^{(s)}=0$ for $s$-wave, $\alpha_\mathbf{k}^{(p_y)}=\arg(\rm{sgn}(\mathbf{k} \cdot \mathbf{n}))$ and $\alpha_\mathbf{k}^{(p\pm ip)}=\arg(k_x \pm ik_y)$ for $p\pm ip$ wave.
$\bm{n}$ is the vector along the nodal direction.  $\varphi$ is the phase differences between the two SCs.
$j$ is the Bragg scattering summation. Within the first BZ of the measured layer, the momentum $\mathbf{k}$ is coupled to three different momenta $\mathbf{k}_j$ in the probe layer, related by Bragg scattering $j$. 
$F_0^{(1)}$ is the part of the free energy that does not depend on $\varphi$. In the last equation, we have taken the zero temperature limit, which gives Eq. \ref{Eq5} in the main text.

For the leading second harmonic term, we have:
\begin{widetext}
\begin{equation}\label{Eq11}
\begin{aligned}
F^{(2)} &=  \frac{1}{4\beta} \mathrm{Tr}[G_0 T]^{4} =  \frac{1}{2\beta} \sum_{j_1,j_2,n,\mathbf{k}} \rm{Tr} [G_b (\mathbf{k},i\omega_n)T_{j_1}G_t(\mathbf{k}_{j_1},i\omega_n)T^\dagger_{j_1} G_b (\mathbf{k}_{j_2},i\omega_n)T_{j_2}G_t(\mathbf{k}_{j_3},i\omega_n)T^\dagger_{j_2} ] \\
&=F_0^{(2)} + \sum_{j_1,j_2,n,\mathbf{k}} \frac{|t|^4 |\Delta_{\mathbf{k}}||\Delta_{\mathbf{k}_{j_2}}||\Delta_{s}|^2 \cos(\alpha_{\mathbf{k}}+\alpha_{\mathbf{k}_{j_2}}+2\varphi)} {\beta(\omega_n^2+|\Delta_{\mathbf{k}}|^2+\xi_{b,\mathbf{k}}^2) (\omega_n^2+|\Delta_s|^2+\xi_{t,\mathbf{k}_{j_1}}^2)(\omega_n^2+|\Delta_{\mathbf{k}_{j_2}}|^2+\xi_{b,\mathbf{k}_{j_2}}^2) (\omega_n^2+|\Delta_s|^2+\xi_{t,\mathbf{k}_{j_3}}^2) },
\end{aligned}
\end{equation}
\end{widetext}
where $j_{1}$ and $j_2$ are Bragg scattering summations. Momentum $\mathbf{k}_{j_3}$ is determined once specify $j_{1}$ and $j_2$ processes.
Here we only kept the phase-dependent second harmonic Josephson couplings. The other terms are included in $F_0^{(2)}$ (phase independent term and also a correction to the first harmonic at fourth order in $t$.). 
%The tunneling scenario is $\mathbf{k}_a$ state in the measured layer is coupling to $\mathbf{k}_{a,i}$ state in the probe layer, which is coupled back to $\mathbf{k}_b$ state in the measured layer. $\mathbf{k}_b$ further connects to $\mathbf{k}_{b,j}$ in the probe layer. 
For the nodal order parameter $p_y$ (odd under mirror), the first harmonic vanishes as seen from Eq. \ref{Eq5} in the main text. 
For the second harmonic here, one possible term in Eq. \ref{Eq11} is $\alpha_{\mathbf{k}_a}=\alpha_{\mathbf{k}_b}=\alpha_{\mathbf{k}}$, which gives the phase $\cos(2\alpha_{\mathbf{k}}+2\varphi)$.
In this case, inserting $\alpha_\mathbf{k}^{(p_y)}=\arg(\rm{sgn}(\mathbf{k} \cdot \mathbf{n}))$ gives a non-vanishing second harmonic coupling.  

The leading third harmonic coupling is generated at sixth order in $t$:
\begin{scriptsize}
\begin{widetext}
\begin{equation}\label{Eq12}
\begin{aligned}
F^{(3)} &=  \frac{1}{6\beta} \mathrm{Tr}[G_0 T]^{6}  = F_0^{(3)} \\
&- \sum_{j_1,j_2,j_3,n,\mathbf{k}} \frac{2 |t|^6 |\Delta_{\mathbf{k}}||\Delta_{\mathbf{k}_{j_2}}||\Delta_{\mathbf{k}_{j_4}}||\Delta_{s}|^3 \cos(\alpha_{\mathbf{k}}+\alpha_{\mathbf{k}_{j_2}}+\alpha_{\mathbf{k}_{j_4}}+3\varphi)} {3\beta{(\omega_n^2+|\Delta_{\mathbf{k}}|^2+\xi_{b,\mathbf{k}}^2) (\omega_n^2+|\Delta_s|^2+\xi_{t,\mathbf{k}_{j_1}}^2)(\omega_n^2+|\Delta_{\mathbf{k}_{j_2}}|^2+\xi_{b,\mathbf{k}_{j_2}}^2) (\omega_n^2+|\Delta_s|^2+\xi_{t,\mathbf{k}_{j_3}}^2)(\omega_n^2+|\Delta_{\mathbf{k}_{j_4}}|^2+\xi_{b,\mathbf{k}_{j_4}}^2) (\omega_n^2+|\Delta_s|^2+\xi_{t,\mathbf{k}_{j_5}}^2) }}.
\end{aligned}
\end{equation}
\end{widetext}
\end{scriptsize}
$j_{1,2,3}$ are Bragg scattering summations. Momentum $\mathbf{k}_{j_{4}}$ and $\mathbf{k}_{j_{5}}$ are determined once specify $j_{1,2,3}$ processes.

\subsection{MATBG-NbSe$_2$ Josephson coupling}
The MATBG layer is described by the continuum model \cite{bistritzer2011moire}:
\begin{align}
&H_{MATBG} = H_t + H_b + H_{tb}, \\
&H_{t} = \sum_{s,\xi,\mathbf{q}} a^\dagger_{t,s,\xi}(\mathbf{q}) \xi \hbar v_f \hat{R}_{+} \mathbf{q} \cdot \bm{\sigma} a_{t,s,\xi}(\mathbf{q}), \\
&H_{b} = \sum_{s,\xi,\mathbf{q}} a^\dagger_{b,s,\xi}(\mathbf{q}) \xi \hbar v_f \hat{R}_{-} \mathbf{q} \cdot \bm{\sigma} a_{b,s,\xi}(\mathbf{q}), \\
&H_{tb} = \sum_{s, \xi, \mathbf{q},\mathbf{q}'}  a^\dagger_{t,s,\xi} (\mathbf{q}) T^{\xi}_{\mathbf{q},\mathbf{q}'} a_{b,s,\xi}(\mathbf{q}').
\end{align}
$a^\dagger_{t/b,s,\xi}$ is a two-component spinor of creation operators for electrons in the two sublattices in the top or bottom (t or b) graphene later, with spin $s$ and valley $\xi$. $\hat{{R}}_{\pm} = \cos{\frac{\theta}{2}} \pm i\sigma_y \sin{\frac{\theta}{2}}$ is the rotation matrix of the top/bottom layer with relative twist angle $\theta$. $v_f$ is the Fermi velocity of the graphene layer. $\sigma^\alpha$ are Pauli matrices that act in sublattice space. $\mathbf{q}$ and $\mathbf{q}'$ are electron momentum related by Bragg scatterings:  $\mathbf{q}-\mathbf{q}'=\{\mathbf{q}_b,\mathbf{q}_{tr},\mathbf{q}_{tl}\}$, where $\mathbf{q}_b = \frac{8\pi\sin{\frac{\theta}{2}}}{3\sqrt{3}a}(-1,0), \mathbf{q}_{tr} = \frac{8\pi\sin{\frac{\theta}{2}}}{3\sqrt{3}a}(\frac{\sqrt{3}}{2},\frac{1}{2}), \mathbf{q}_{tl} = \frac{8\pi\sin{\frac{\theta}{2}}}{3\sqrt{3}a}(-\frac{\sqrt{3}}{2},\frac{1}{2})$. $a$ is the bond length in graphene. $T^{\xi}_{\mathbf{q},\mathbf{q}'}$ is the interlayer tunneling matrix, given in Ref. \cite{bistritzer2011moire}.

NbSe$_2$ is modelled by a tight-binding model with three Nb $d$ orbitals: $d_{z^2,\uparrow/\downarrow}$, $d_{xy,\uparrow/\downarrow}$, and $d_{x^2-y^2,\uparrow/\downarrow}$.
The Slater-Koster hopping between these orbitals is given in Ref. \cite{liu2013three}.

The Josephson current of the twisted interface is calculated by including the pairing potential in each layer and also the momentum-resolved interlayer tunneling.
The pairing potential is added in the BCS mean-field way. For NbSe$_2$, we use a constant superconducting gap $\Delta_s=0.67$ meV. 
For MATBG, we assume the nodal and chiral order parameter takes the form:
\begin{align}
    &\Delta_{p_x} = \Delta_0 \cos(\alpha_\mathbf{k}),\\
    &\Delta_{p_y} = \Delta_0 \sin(\alpha_\mathbf{k}),\\
    &\Delta_{p\pm ip} = \Delta_0 e^{i\alpha_\mathbf{k}},
\end{align}
with $\Delta_0=0.2$ meV. $\alpha_\mathbf{k}$ is the angle between $\mathbf{k}$ and the $x$ axis, measured relative to the center of mini-BZ.
% In reality, the order parameters can have varying amplitude in momentum space. This affects the quantitative Josephson current but does not change the results qualitatively.

The Cooper pair tunneling events include Bragg scatterings within the range of the first BZ of the graphene layer. 
% For example, for each Dirac fermion $h_t(\mathbf{k})$ in the top layer of the MATBG, it connects to three states $h_{NbSe_2}(\mathbf{k}_{1,2,3})$ with different momenta:
% \begin{align}
%     \begin{psmallmatrix}
%     h_t(\mathbf{k}) & T_1(\mathbf{k},\mathbf{k}_1) & T_2(\mathbf{k},\mathbf{k}_2) & T_3(\mathbf{k},\mathbf{k}_3)  \\
%     T^\dagger_1(\mathbf{k},\mathbf{k}_1) & h_{NbSe_2}(\mathbf{k}_1) & 0 & 0 \\
%     T^\dagger_2(\mathbf{k},\mathbf{k}_2) & 0 & h_{NbSe_2}(\mathbf{k}_2) & 0  \\
%     T^\dagger_3(\mathbf{k},\mathbf{k}_3) & 0 & 0 & h_{NbSe_2}(\mathbf{k}_3) \\
%     \end{psmallmatrix}.
% \end{align}
Including the microscopic orbital symmetry, the largest interlayer coupling comes from the $p_z$ orbital of graphene and $d_{z^2}$ orbital of NbSe$_2$. The interlayer tunneling term is:
\begin{align}
    &H_{inter} = \sum_{s,\xi,\mathbf{k},j,\sigma} a_{t,s,\xi,\sigma}^\dagger  (\mathbf{k}) T^{\sigma}_{j}  d_{{z^2}, s} (\mathbf{k}'), \\
    &T^{\sigma}_{j} = t  e^{i\mathbf{G}_{T,j}\bm{\tau}_{\sigma}}.
\end{align}
${\sigma}$ is the sublattice index, with the corresponding sublattice vector $\bm{\tau}_{\sigma}$.
$j$ is the Bragg scattering processes, which relates momentum by:
\begin{align}
    \mathbf{k} + \mathbf{K_{g,\xi}} + \mathbf{G}_{j,T} = \mathbf{k}'+\mathbf{G}_{j,\mathrm{NbSe}_2}.
\end{align}
$\mathbf{G}_{j,T}$ and $ \mathbf{G}_{j,\mathrm{NbSe}_2}$ are reciprocal lattice vectors in the top graphene layer and NbSe$_2$, respectively. $\mathbf{K_{g,\xi}}$ is the $K$ point of the top graphene.
% For the calculations in the main text, we take the value $t=0.1$meV. 

To calculate the Zeeman field dependent Josephson coupling, we include Ising and Rashba SOC in the MATBG continuum model, with the SOC strength $\lambda_{\mathrm{Ising}}$ and $\lambda_{\mathrm{Rashba}}=1$ meV, consistent with the value reported in the literature \cite{arora2020superconductivity}. Here, in this twisted Josephson junction setup, both the inserted WSe$_2$ tunneling barrier and NbSe$_2$ can generate SOC in MATBG. 

\subsection{Interlayer tunneling dependence of the Josephson coupling}
\begin{figure}[h]
    \centering
    \includegraphics[width=\linewidth]{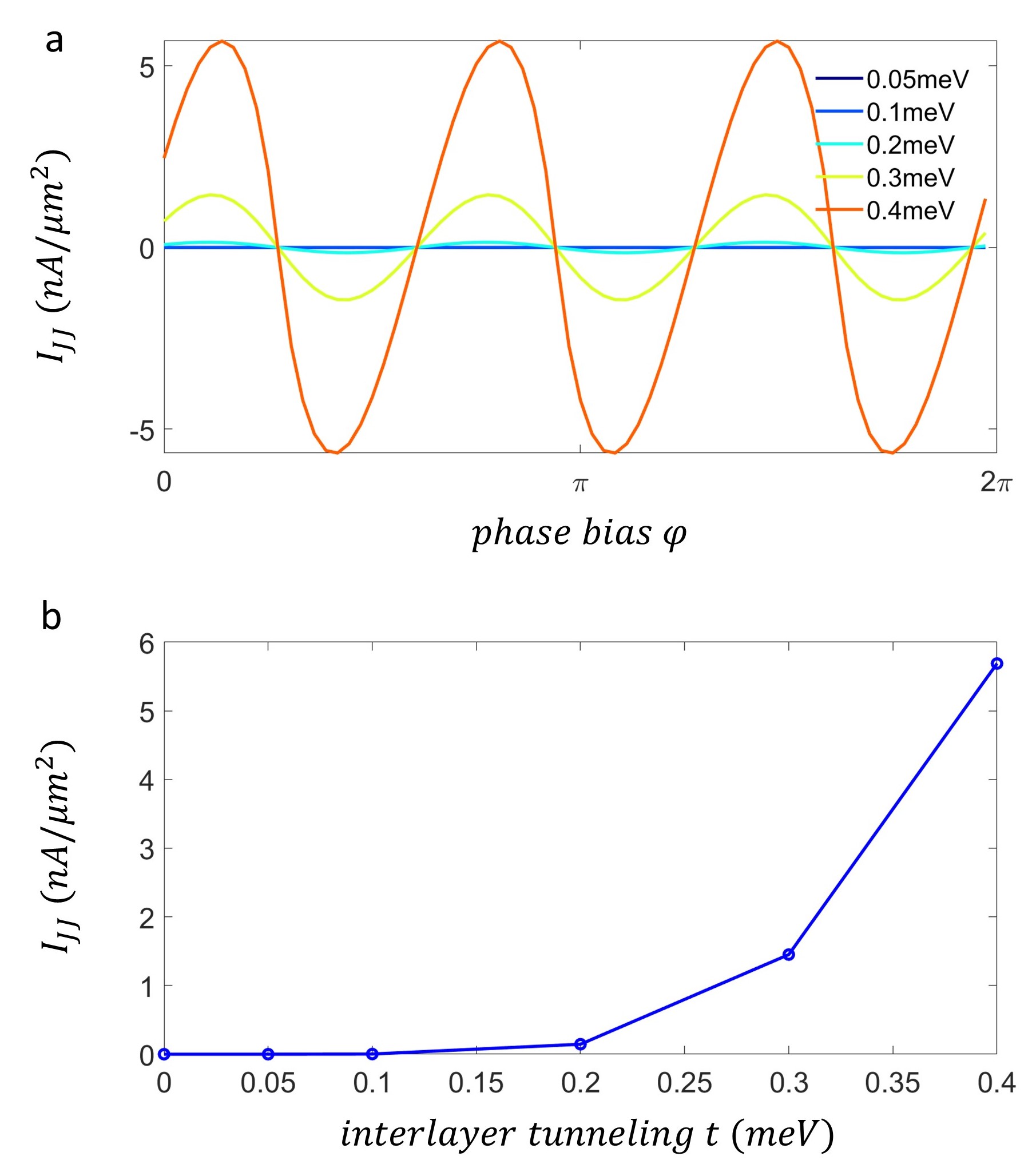}
    \caption{\textbf{Critical current dependence on the tunneling amplitude $t$ in the case of a $p\pm ip $ order parameter in twisted NbSe$_2$-MATBG junction at $\theta=19.5^\circ$} 
    (a) The current phase relation for different interlayer tunnleing amplitude $t$.
    (b) The critical current versus $t$.}
    \label{figure6}
\end{figure}
As discussed in the main text, the critical current of nontrivial order parameter can be enhanced by increasing the interlayer tunneling strength $t$. Fig. \ref{figure6} shows the critical current for the $p+ip$ case of the NbSe$_2$-MATBG twisted junction at twist angle $\theta=19.5^\circ$. $I_c$ is significantly increased to $\sim 5$ $\rm{nA}/\mu m^2$ when $t=0.4$ meV. When increasing $t$ from $0.05$ meV to $0.3$ meV, $I_c$ scales approximately as $t^6$, as expected.

\subsection{Suppression of interlayer hybridization by tunneling barrier}
\begin{figure}
    \centering
    \includegraphics[width=\linewidth]{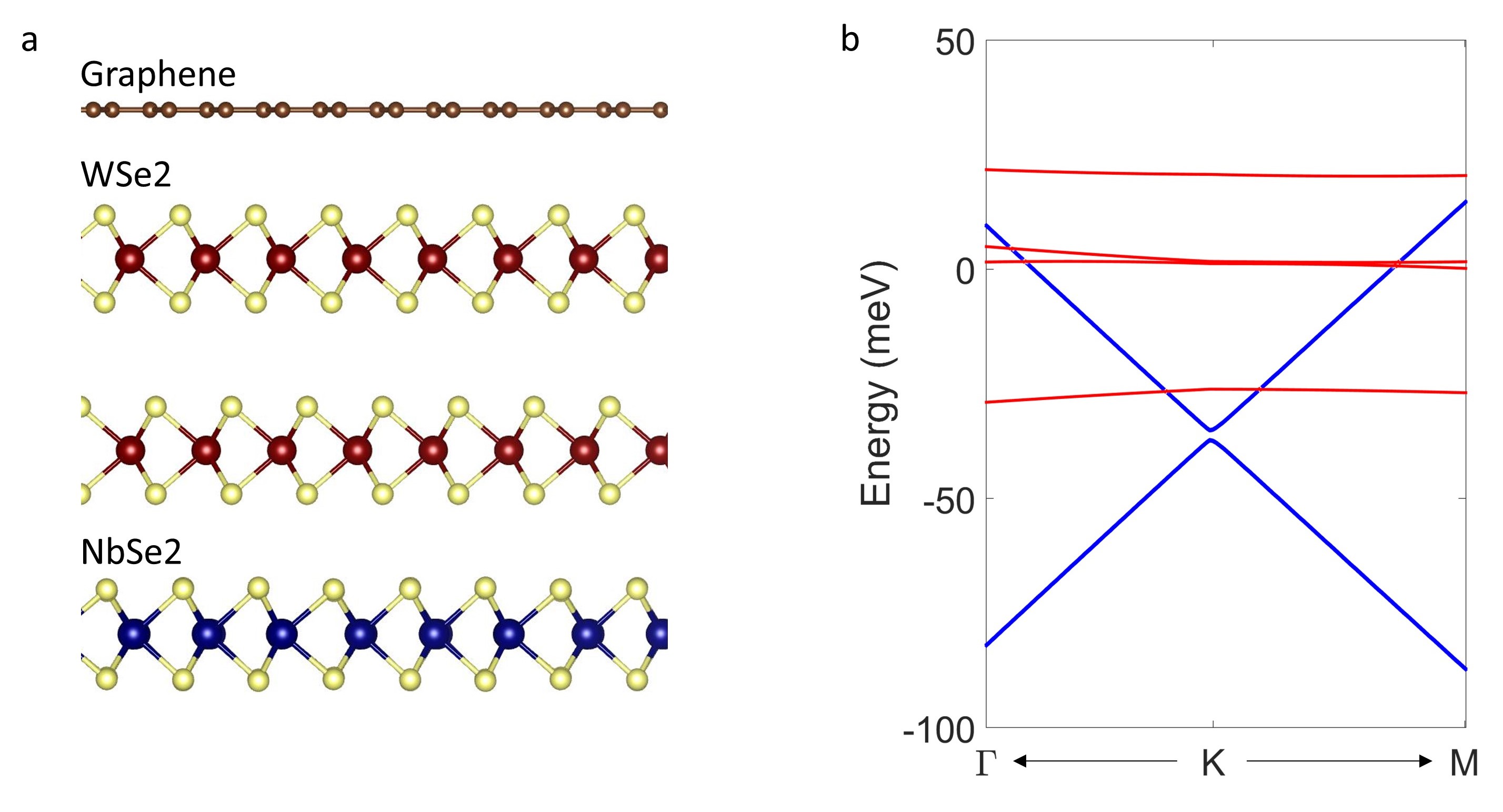}
    \caption{
    (a) The graphene-bilayer WSe$_2$-monolayer NbSe$_2$ heterostructure.
    (b) Band structure of the heterostructure. Red and blue colors represent the weight of the wave function on NbSe$_2$ (red) and graphene (blue). The WSe$_2$ bands are outside of the energy window shown here. }
    \label{figure7}
\end{figure}
We performed the density function theory (DFT) calculation using the Vienna Ab initio Simulation Package (VASP) \cite{kresse1996efficiency,kresse1996efficient}. 
 The exchange-correlation is described by the Perder-Burke-Ernzerhof (PBE) formulation under the generalized gradient approximation (GGA) \cite{perdew1996generalized}.
Here we consider a heterostructure of graphene-bilayer WSe$_2$-monolayer NbSe$_2$, as shown in Fig. \ref{figure7}a. The band structure is shown in Fig. 7b. The red and blue color represents the weight of wave function on NbSe$_2$ (red) and graphene (blue).
In the low energy region, we see that the graphene Dirac cone approximately retains its original shape and is shifted by $35$ meV, due to the work function differences, which can be gated to the charge neutrality. For the case of a few layers of graphene (such as MATBG), we expect the work function differences to be of a similar magnitude. 
% The inserted WSe$_2$ tunneling barrier suppresses the hybridization, resulting in the graphene Dirac cone.
% the estimated level anti-crossing (black circle) is around $t\approx 2meV$. In these regions, we can see the mixed color of red and blue, which shows the interlayer hybridization. If increasing the WSe2 layer thickness, $t$ can be further suppressed.

\newpage
\subsection{Magnetic field dependent Josephson couplings}
\subsubsection{Zeeman effect}
Here we add Zeeman field, Rashba and Ising SOC in the toy model to study the Zeeman field induced Josephson couplings. The toy model Hamiltonian is:
\begin{widetext}
\begin{equation}
    \begin{aligned}
        H_t = \frac{1}{2}\sum_\mathbf{k}\Psi^\dag_{t,\mathbf{k}} \begin{pmatrix}
    \xi_t(\mathbf{k},\sigma)+g_t\mathbf{B}\cdot \bm{\sigma} +\lambda_t \sigma_z & i\sigma_y\Delta_t(\mathbf{k})\\
    -i\sigma_y\Delta_t^*(\mathbf{k}) & -(\xi_t(-\mathbf{k},\sigma)+g_t\mathbf{B}\cdot \bm{\sigma}-\lambda_t \sigma_z)^T 
    \end{pmatrix}\Psi_{t,\mathbf{k}},
    \end{aligned}
\end{equation}
\begin{equation}
    \begin{aligned}
        H_b = \frac{1}{2}\sum_\mathbf{k}\Psi^\dag_{b,\mathbf{k}} \begin{pmatrix}
    \xi_b(\mathbf{k},\sigma)+g_b\mathbf{B}\cdot \bm{\sigma}+\lambda_b \mathbf{k}\times \bm{\sigma} & i\sigma_y\Delta_b(\mathbf{k})\\
    -i\sigma_y\Delta_b^*(\mathbf{k}) & -(\xi_b(-\mathbf{k},\sigma)+g_b\mathbf{B}\cdot \bm{\sigma}-\lambda_b \mathbf{k}\times \bm{\sigma})^T 
    \end{pmatrix}\Psi_{b,\mathbf{k}},
    \end{aligned}
\end{equation}
\end{widetext}
where $H_t$ is the probe layer and $H_b$ is the measured layer. $\Psi^\dag_{l,\mathbf{k}}=(c_{l,\mathbf{k},\sigma}^\dag, c_{l,\mathbf{k},\Bar{\sigma}}^\dag, c_{l,\mathbf{-k},\sigma}, c_{l,\mathbf{-k},\Bar{\sigma}})$, and $c_{l,\mathbf{k},\sigma}$ annihilates a state with spin $\sigma$ and momentum $\mathbf{k}$ in layer $l$. The momentum $\mathbf{k}$ is measured relative to the $K$ point. The Ising SOC $\lambda_t \sigma_z$ is included in the probe layer, which mimics the strong Ising SOC in NbSe$_2$. The interlayer tunneling remains the form of Eq. 4 in the main text. 
For simplicity, we do not consider Bragg scatterings in the analytical expression shown below and momentum $\mathbf{k}$ determines the $\mathbf{k}'$ by the relation $\mathbf{k}+\mathbf{K}_b = R(\theta) (\mathbf{k}'+\mathbf{K}_t)$. 
The second order perturbation gives:
\begin{widetext}
\begin{equation}
\begin{aligned}
    F^{(1)} &=  \frac{1}{\beta} \sum_{n,\mathbf{k}} \rm{Tr} [G_b (\mathbf{k},i\omega_n)TG_t(\mathbf{k}',i\omega_n)T^\dagger ] \\
    &= \sum_{\mathbf{k},n=0}^{n=\infty} 
    8 \Delta_b\Delta_t \Re \Bigg[  
    \frac{\cos({\varphi}+\alpha_{\mathbf{k}})}{\beta f_1f_2}\Big[ -(-g_t^2|\mathbf{B}|^2+\Delta_t^2+\xi_t^2+\lambda_t^2+\omega_n^2)
     (-g_b^2|\mathbf{B}|^2+\Delta_b^2+\xi_b^2+\lambda_b^2|\mathbf{k}|^2+\omega_n^2)  \\
    & + 4g_t\omega_n (\omega_n g_b|\mathbf{B}|^2 - i\xi_b \lambda_b (\mathbf{k} \times \mathbf{B}) ) 
    \Big]  +\frac{4\sin{(\varphi+\alpha_{\mathbf{k}})}}{\beta f_1f_2} \Big[
    \lambda_t\lambda_b (\mathbf{k} \cdot \mathbf{B}) (\xi_t g_b + \xi_b g_t )
    \Big]
    \Bigg],
    \end{aligned}
\end{equation}
where,
\begin{equation}
    \begin{aligned}
    f_1 &= 
    (g_t|\mathbf{B}|^2 + \omega_n^2 + \lambda_t^2 - \Delta_t^2 - \xi_t^2)^2 + 4(\Delta_t^2\omega_n^2 + \Delta_t^2 \lambda_t^2 + \omega_n^2 \xi_t^2), \\
    f_2 &=
    g_b^4|\mathbf{B}|^4 + 2g_b^2|\mathbf{B}|^2(\omega_n^2-\Delta_b^2-\xi_b^2) + (\Delta_b^2+\omega_n^2+\xi_b^2+\lambda_b^2|\mathbf{k}|^2)^2 - 4\lambda_b^2|\mathbf{k}|^2\xi_b^2 \\
    &+ 2g_b^2\lambda_b^2 ((\mathbf{k}\cdot \mathbf{B})^2-(\mathbf{k}\times \mathbf{B})^2) + 8i\omega_n \xi_b \lambda_b g_b (\mathbf{k}\times \mathbf{B}).
    \end{aligned}
\end{equation}
We sum the Matsubara frequency in pairs: $\omega_{n} = (2n+1)/\beta$ and $\omega_{-n-1} = -(2n+1)/\beta$.
To the lowest order in $|B|$, the above equation (28) can be written in the form:
\begin{equation}
\begin{aligned}
    F^{(2)} 
    &= \sum_{\mathbf{k}} \Big [F_{0\mathbf{k}} + F_{1\mathbf{k}}[(\mathbf{k}\cdot \mathbf{B})^2,(\mathbf{k}\times \mathbf{B})^2,|\mathbf{B}|^2] \Big]\cos(\varphi + \alpha_{\mathbf{k}}+\beta_{\mathbf{k}, \mathbf{B}}),
    \end{aligned}
\end{equation}
\end{widetext}
where $F_{0\mathbf{k}}$ is independent of the Zeeman field and has $C_3$ symmetry. $F_{1\mathbf{k}}$ is a function of $(\mathbf{k}\cdot \mathbf{B})^2$,$(\mathbf{k}\cdot\mathbf{B})^2$, $|\mathbf{B}|^2$, second order in $|\mathbf{B}|$. There is also a phase shift $\beta_{\mathbf{k}, \mathbf{B}} = b_{\mathbf{k}} (\mathbf{k} \cdot \mathbf{B}) + \mathcal{O}(|\mathbf{B}|^2)$, where $b_\mathbf{k}$ (obtained from combining the sine and the cosine in Eq. 27) depends on microscopic parameters, e.g., momentum $\mathbf{k}$, $g_{t/b}$ and $\lambda_{t/b}$. 

For the $s-$wave case, we have a constant $\alpha_{\mathbf{k}}$, which gives a non-zero $F^{(2)} = \sum_{\mathbf{k}} F_{0\mathbf{k}} \cos({\varphi})$ in the absence of Zeeman field. Given the existing first order term, both the $\mathbf{B}$ induced $F_{1\mathbf{k}}$ term and the phase shift $\beta_{\mathbf{k},\mathbf{B}}$ give a $|\mathbf{B}|^2$ dependent critical current. 

For the $p+ip$ case, $\sum_{\mathbf{k}} F_{0\mathbf{k}} \cos({\varphi+\alpha_{\mathbf{k}}})$ vanishes by $C_3$ symmetry.
In the presence of an in-plane field, $\beta_{\mathbf{k},\mathbf{B}} = b_{\mathbf{k}} (\mathbf{k} \cdot \mathbf{B})$ destroys the destructive interference in the summation $\sum_{\mathbf{k}} F_{0\mathbf{k}} \cos({\varphi+\alpha_{\mathbf{k}}+\beta_{\mathbf{k},\mathbf{B}}})$. As a result, a first-harmonic Josephson coupling with a magnitude proportional to $|\mathbf{B}|$ is generated (seen directly from Eq. 27).
% \begin{widetext}
% \begin{equation*}
% \begin{aligned}
% \sum_{\mathbf{k}\in BZ} F_{0\mathbf{k}} \cos({\varphi+\alpha_{\mathbf{k}}+\beta_{\mathbf{k},\mathbf{B}}}) &= \sum_{\substack{\mathbf{k}\in reduced BZ, \\ \ n=\{0,1,2\}}} F_{0\mathbf{k}} \cos({\varphi+(\alpha_{\mathbf{k}}+n\frac{2\pi}{3}})+\beta_{\mathbf{k},\mathbf{B},n}) \\ 
% &\approx \sum_{\mathbf{k} \in reduced BZ} F_{0\mathbf{k}}[\frac{\alpha_{\mathbf{B},2}+\alpha_{\mathbf{B},3}}{2}-\alpha_{\mathbf{B},1}]\sin(\varphi+\alpha_{\mathbf{k}})
% +F_{0\mathbf{k}}\frac{\sqrt{3}}{2} [\alpha_{\mathbf{B},3}-\alpha_{\mathbf{B},2}]\cos(\varphi+\alpha_{\mathbf{k}}) \\
% &=|\mathbf{B}|\sum_{\mathbf{k} \in reduced BZ} F_{0\mathbf{k}}[\frac{c_2+c_3}{2}-c_1]\sin(\varphi+\alpha_{\mathbf{k}})
% +F_{0\mathbf{k}}\frac{\sqrt{3}}{2} [c_3-c_2]\cos(\varphi+\alpha_{\mathbf{k}})
% \end{aligned}
% \end{equation*}
% \end{widetext}
% Where we expressed the phase shift $\alpha^n_{\mathbf{k},\mathbf{B}} = \alpha_{\mathbf{B},R^{n-1}_3 \mathbf{k}}$, where $R_3$ is a rotation matrix by $2\pi/3$, linear in the Zeeman field strength.

\subsubsection{Orbital effect}
Here we derive the orbital effect of an in-plane magnetic field. 
If the tunneling between the two superconductors is perfectly momentum-conserving, then any arbitrarily small in-plane orbital field completely decouples the two order parameters. To mimic the effect of the finite size of the system and the effect of disorder, we first relax the momentum conservation assumption made in Eq. 4, and instead, we write the tunneling element in the absence of magnetic field as $t(\mathbf{p},\mathbf{p}')=t_{(\frac{\mathbf{p}+\mathbf{p}'}{2})} \cdot f(\mathbf{p}-\mathbf{p}')$. $f$ is a real and symmetric function peaked at zero (taking $f$ to be a delta function recovers the momentum-conserving limit).
We assume an in-plane magnetic field, and write the vector potential as $\mathbf{A} = z\mathbf{B}\times\Hat{\mathbf{z}}$. The Hamiltonian is written as:
\begin{equation}
\mathcal{H}=\mathcal{H}_t+\mathcal{H}_b+{T},
\end{equation}
where $\mathcal{H}_{t,b}$ are as defined in Eq. 3, and $T$ is given by
\begin{widetext}
\begin{equation}
\begin{aligned}
T=\sum_{p,p'}T_{\mathbf{p},\mathbf{p'}}=\sum_{p,p'}\sum_{G_1,G_2}t_{(\frac{\Tilde{p}+\Tilde{p}'}{2})}\Psi^\dag_{t,\mathbf{p}} \begin{pmatrix}
    f(\Tilde{\mathbf{p}}-\Tilde{\mathbf{p}}'+\mathbf{q}) & 0\\
    0 & -f(\Tilde{\mathbf{p}}'-\Tilde{\mathbf{p}}+\mathbf{q}) 
    \end{pmatrix}\Psi_{b,\mathbf{p'}},
\end{aligned}
\end{equation}
\end{widetext}
where $\Tilde{\mathbf{p}}=\mathbf{p}+G_1$, $\Tilde{\mathbf{p}}'=\mathbf{p}'+G_2$ and $\mathbf{q}=\frac{e}{\hbar}d\mathbf{B}\times \hat{\mathbf{z}} 
$ is the momentum boost due to the magnetic field. $d$ is the distance between the two SCs. Using the fact that $f$ is symmetric we can expand to first order:
\begin{equation}
T_{\mathbf{p},\mathbf{p'}}=\sum_{G_1,G_2}t_{(\frac{\Tilde{p}+\Tilde{p}'}{2})}\Psi^\dag_{t,\mathbf{p}} 
    (f(\Tilde{\mathbf{p}}-\Tilde{\mathbf{p}}')\tau_z +\mathbf{q}\cdot\nabla f\tau_0)
    \Psi_{b,\mathbf{p'}}.
\end{equation}
The $n^{th}$ contribution to the free energy is given by:
\begin{widetext}
\begin{equation}
    F^{(n)} = \frac{1}{n\beta}\sum_{\{ \mathbf{p}\},\{ \mathbf{p}'\},\omega}\mathrm{Tr}[\prod_{i=1}^n (T_{\mathbf{p}_{i},\mathbf{p}'_{i-1}} G_t(\mathbf{p}_{i},\omega) T_{\mathbf{p}_{i},\mathbf{p}'_{i}} G_b(\mathbf{p}'_{i},\omega))].
\end{equation}
Using Eq.32 and the trace invariance to circular shifts we can show that its linear expansion in $\mathbf{q}$ is given by:
\begin{equation}
\begin{aligned}
    F^{(n)} =F^{(n)}(\mathbf{q}=0) 
    +\frac{1}{\beta}\sum_{\substack{\{ \mathbf{p}\},\{ \mathbf{p}'\},\omega, \\ G_1,G_2}}
    \mathrm{Tr}\left[ \left( t_{(\frac{\Tilde{p}_1+\Tilde{p}'_n}{2})}\mathbf{q}\cdot\nabla f(\Tilde{\mathbf{p}}_{1}-\Tilde{\mathbf{p}}'_{n})G_t(\mathbf{p}_{1},\omega) T^{q=0}_{\mathbf{p}_{1},\mathbf{p}'_{1}} G_b(\mathbf{p}'_{1},\omega)+ \right. \right. \\  \left.  \left. 
 T^{q=0}_{\mathbf{p}_{1},\mathbf{p}'_{n}}G_t(\mathbf{p}_{1},\omega)  t_{(\frac{\Tilde{p}_1+\Tilde{p}'_1}{2})}\mathbf{q}\cdot\nabla f(\Tilde{\mathbf{p}}_{1}-\Tilde{\mathbf{p}}'_{1})G_b(\mathbf{p}'_{1},\omega) \right)\prod_{i=2}^n \left(T^{q=0}_{\mathbf{p}_{i},\mathbf{p}'_{i-1}} G_t(\mathbf{p}_{i},\omega) T^{q=0}_{\mathbf{p}_{i},\mathbf{p}'_{i}} G_b(\mathbf{p}'_{i},\omega)\right)\right].
\end{aligned}
\end{equation}
Since we sum over all momenta and Matsubara frequency, we can use the following map for the second term in the parenthesis: $\omega\leftrightarrow-\omega$, $\mathbf{p}'_{i+1}\leftrightarrow\mathbf{p}'_{n-i}$ and $\mathbf{p}_{i+2}\leftrightarrow\mathbf{p}_{n-i}$ ($\mathbf{p}_1$ maps to itself) to get
\begin{equation}
\begin{aligned}
    \nabla_{\mathbf{q}}F^{(n)} =\frac{1}{\beta}\sum_{\substack{\{ \mathbf{p}\},\{ \mathbf{p}'\},\omega, \\ G_1,G_2}}t_{(\frac{\Tilde{p}_1+\Tilde{p}'_n}{2})}\nabla f(\Tilde{\mathbf{p}}_{1}-\Tilde{\mathbf{p}}'_{n})
    \mathrm{Tr}\left[ G_t(\mathbf{p}_{1},\omega) T^{q=0}_{\mathbf{p}_{1},\mathbf{p}'_{1}} G_b(\mathbf{p}'_{1},\omega)\prod_{i=2}^n \left(T^{q=0}_{\mathbf{p}_{i},\mathbf{p}'_{i-1}} G_t(\mathbf{p}_{i},\omega) T^{q=0}_{\mathbf{p}_{i},\mathbf{p}'_{i}} G_b(\mathbf{p}'_{i},\omega)\right) +  \right. \\    \left. 
 T^{q=0}_{\mathbf{p}_{1},\mathbf{p}'_{1}}G_t(\mathbf{p}_{1},-\omega)  G_b(\mathbf{p}'_{n},-\omega)\prod_{i=n}^2 \left(T^{q=0}_{\mathbf{p}_{i},\mathbf{p}'_{i}} G_t(\mathbf{p}_{i},-\omega) T^{q=0}_{\mathbf{p}_{i},\mathbf{p}'_{i-1}} G_b(\mathbf{p}'_{i-1},-\omega)\right) \right].
\end{aligned}
\end{equation}
Performing a circular shift for the second term, and using the fact that $G_l^\dagger(\mathbf{p},\omega)=G_l(\mathbf{p},-\omega)$ let us write it as:
\begin{equation}
    \nabla_{\mathbf{q}}F^{(n)} =\frac{1}{\beta}\sum_{\substack{\{ \mathbf{p}\},\{ \mathbf{p}'\},\omega, \\ G_1,G_2}}t_{(\frac{\Tilde{p}_1+\Tilde{p}'_n}{2})}\nabla f(\Tilde{\mathbf{p}}_{1}-\Tilde{\mathbf{p}}'_{n})
    \mathrm{Tr}\left[ M +  M^\dagger \right],
\end{equation}
where:
\begin{equation}
    M=G_t(\mathbf{p}_{1},\omega) T^{q=0}_{\mathbf{p}_{1},\mathbf{p}'_{1}} G_b(\mathbf{p}'_{1},\omega)\prod_{i=2}^n \left(T^{q=0}_{\mathbf{p}_{i},\mathbf{p}'_{i-1}} G_t(\mathbf{p}_{i},\omega) T^{q=0}_{\mathbf{p}_{i},\mathbf{p}'_{i}} G_b(\mathbf{p}'_{i},\omega)\right).
\end{equation}
\end{widetext}
Both $T^{q=0}_{\mathbf{p},\mathbf{p}'}$ and $G_i(\mathbf{p},\omega)$ are traceless and can be written as a sum of Pauli matrices in Nambu space. Hence, $M$ is given by a linear sum of products of an odd number of Pauli matrices. After summing over Matsubara frequencies, tracing over $M$ gives a purely imaginary contribution. Using the fact that $\mathrm{Tr}\left[M^\dagger \right] = \mathrm{Tr}\left[ M\right]^*$ we find that $\nabla_{\mathbf{q}}F^{(n)}=0$.

\subsubsection{In-plane current effect}
A similar effect to that of an in-plane field can be obtained by driving an in-plane current through one of the SCs. In this case the order parameter acquires a finite momentum $\Delta_t(\mathbf{p})\rightarrow\Delta_t(\mathbf{p}) e^{i2\mathbf{q}\cdot \mathbf{R}}$ (Assuming current in the top layer, where $\mathbf{R}$ is the center of mass coordinate). This modifies the BdG Hamiltonian of the top layer.
% Assuming current in the top layer, the Hamiltonian is modified as:
% \begin{equation}
% \begin{aligned}
%     H_t 
%     &= \sum_{\mathbf{k},\sigma} \xi_t (\mathbf{k}) c^\dagger_{t,\mathbf{k},\sigma} c_{t,\mathbf{k},\sigma} \\
%     &+ \sum_{\mathbf{k},\sigma} (\Delta_{t,\mathbf{k}}^* c_{t,\mathbf{k}+\bm{q},\sigma}  c_{t,-\mathbf{k}-\bm{q},\sigma'} + c.c.) 
%     \end{aligned}
% \end{equation}
The bottom Hamiltonian is left unchanged, and the inter-layer tunneling $T_{\mathbf{p},\mathbf{p'}}$ is modified in the same way as in the previous section, allowing for non momentum-conserving tunneling between the two SCs. To lowest order in $\mathbf{q}=0$, we can set $\mathbf{q}=0$ in the tunneling matrix element (the linear in $\mathbf{q}=0$ term from the matrix element was shown to vanish in the previous section). The Hamiltonian is given by:
\begin{widetext}
\begin{align}
    \mathcal{H}&=\mathcal{H}_t+\mathcal{H}_b+{T}, \\
    \mathcal{H}_t&=\frac{1}{2}\sum_\mathbf{p} \Psi^\dag_{t,\mathbf{p}} \begin{pmatrix}
    \xi_t(\mathbf{p}+\mathbf{q},\sigma) & i\sigma_y\Delta_t(\mathbf{p})\\
    -i\sigma_y\Delta_t^*(\mathbf{p}) & -(\xi_t^T(-\mathbf{p}+\mathbf{q},\sigma)) 
    \end{pmatrix}\Psi_{t,\mathbf{p}}, \\
    \mathcal{H}_b&=\frac{1}{2}\sum_\mathbf{p} \Psi^\dag_{b,\mathbf{p}} \begin{pmatrix}
    \xi_b(\mathbf{p},\sigma) & i\sigma_y\Delta_b(\mathbf{p})\\
    -i\sigma_y\Delta_b^*(\mathbf{p}) & -(\xi_b^T(-\mathbf{p},\sigma)) 
    \end{pmatrix}\Psi_{b,\mathbf{p}}, 
\end{align}
where $\Psi^\dag_{t,\mathbf{p}}=(c_{t,\mathbf{p}+\mathbf{q},\sigma}^\dag, c_{t,\mathbf{p}+\mathbf{q},\Bar{\sigma}}^\dag, c_{t,-\mathbf{p}+\mathbf{q},\sigma}, c_{t,-\mathbf{p}+\mathbf{q},\Bar{\sigma}})$, $\Psi^\dag_{b,\mathbf{p}}=(c_{b,\mathbf{p},\sigma}^\dag, c_{b,\mathbf{p},\Bar{\sigma}}^\dag, c_{b,\mathbf{-p},\sigma}, c_{b,\mathbf{-p},\Bar{\sigma}})$, $\mathbf{p}'$ is defined as in Eq. 4, and $T$ is as defined in Eq. 32, substituting $\mathbf{q}=0$. 
Expanding the free energy to second order in $T$ and to first order in $\mathbf{q}$ gives:
\begin{equation}
\begin{aligned}
F^{(1)} &= \frac{1}{\beta} \sum_{\mathbf{p},\mathbf{p}',\omega} \mathrm{Tr}[G_b(\mathbf{p},\omega) T_{\mathbf{p},\mathbf{p}'} G_t (\mathbf{p}',\omega) T_{\mathbf{p},\mathbf{p}'}]  \\
&= \sum_{\substack{\{ \mathbf{p}\},\{ \mathbf{p}'\},\omega, \\ G_1,G_2}} \frac{ 4(t_{(\frac{\Tilde{p}+\Tilde{p}'}{2})}f(\Tilde{\mathbf{p}}-\Tilde{\mathbf{p}}'))^2 |\Delta_{s}| |\Delta_{\mathbf{p}'}| \cos{(\varphi+\alpha_{\mathbf{p}'})}}{\beta(|\Delta_{\mathbf{p}'}|^2+\omega_{n}^2+\xi^2_{b,\mathbf{p}'})(E_{t,\mathbf{p}}+\mathbf{\nabla}\xi_t (\mathbf{p})\cdot\mathbf{q}-i\omega)(E_{t,\mathbf{p}}-\mathbf{\nabla}\xi_t (\mathbf{p})\cdot\mathbf{q}+i\omega)}.
%\\ &\stackrel{(T=0)}{=} \sum_{i,\mathbf{k}} \frac{t^2|\Delta_{s}| |\Delta_{\mathbf{k}}| \cos{(\varphi+\alpha_{\mathbf{k}})}}{16E_{b,\mathbf{k}}E_{t,\mathbf{k}_i} (E_{b,\mathbf{k}} + E_{t,\mathbf{k}_i})}.
\end{aligned}
\end{equation}
The linear order term in $\mathbf{q}$ vanishes. 
\end{widetext}

\begin{figure}[H]
    \centering
    \includegraphics[width=\linewidth]{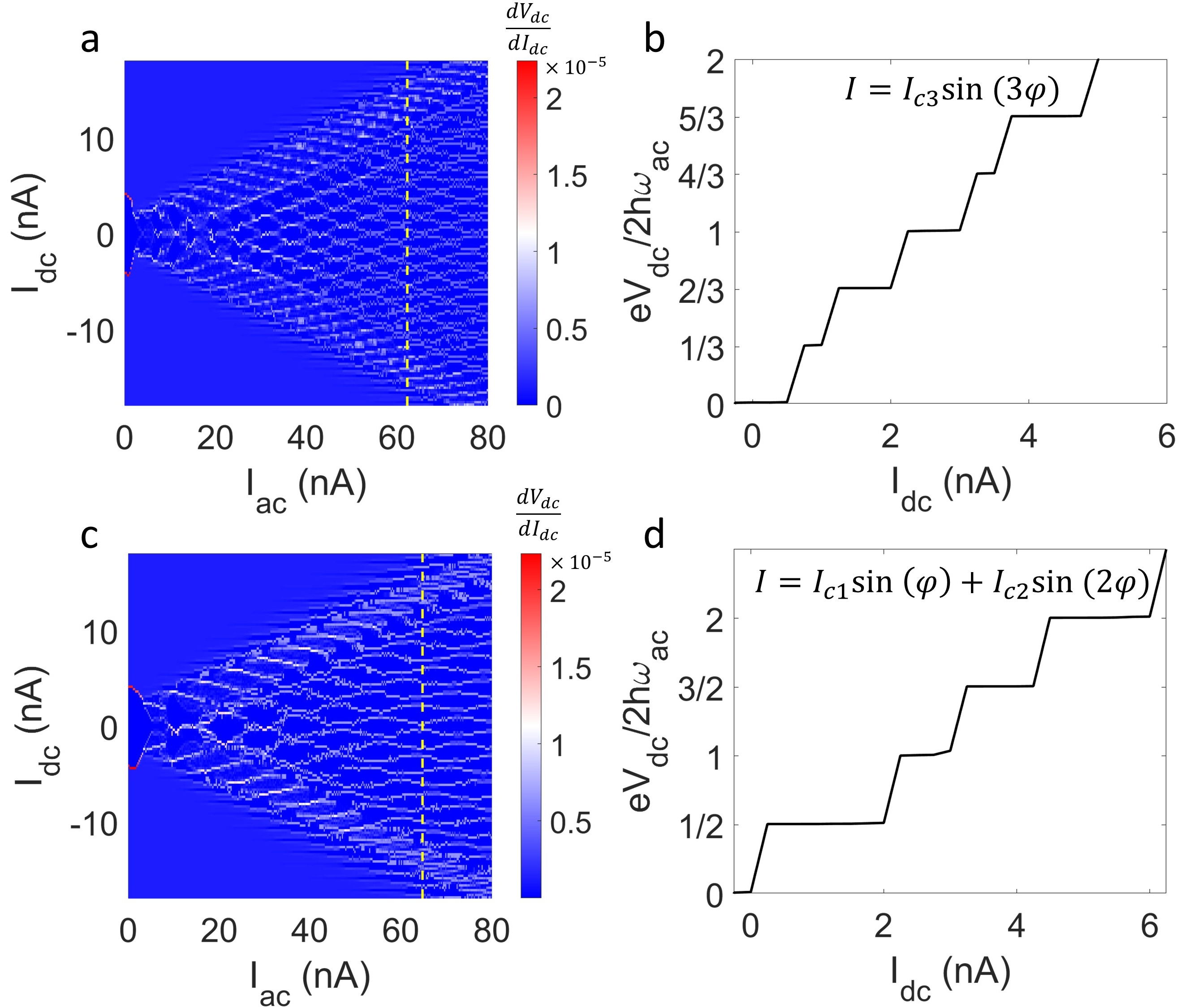}
    \caption{\textbf{Fractional Shapiro steps} 
    (a) The $\frac{dV_{dc}}{dI_{dc}}$ signal versus $I_{ac}$ and $I_{dc}$ parameters in the current bias, with the current-phase relation $I (\varphi) = I_{c3}\sin{3\varphi}$, $I_{c3} = 5$nA. The parameters used in this calculation are given in the text.
    (b) $V_{dc}$ as a function of $I_{dc}$ at $I_{ac}=65$nA for the current-phase relation used in (a).  
    (c) Same as (a) for the current-phase relation $I (\varphi) = I_{c1}\sin{\varphi} + I_{c2}\sin{2\varphi}$, $I_{c1} = 2\mathrm{nA}, I_{c2} = 4$nA.
    (d) $V_{dc}$ as a function of $I_{dc}$ at $I_{ac}=65$nA for the current-phase relation used in (c). }
    \label{figure8}
\end{figure}

\subsection{Fractional Shapiro steps}

To calculate the Shaprio steps, we use the RCSJ model \cite{barone1982physics}. The Josephson junction is described by a circuit composed of a Josephson element, resistor and capacitor in parallel. This model gives the Josephson dynamics under microwave irradiation. Assuming the junction is current biased, we have:
\begin{equation}
\begin{aligned}
&I_{bias}  = I_{dc} + I_{ac} \cos{(\omega t)} = I_{JJ} (\varphi) + \frac{V_{JJ}}{R} + C \frac{dV_{JJ}}{dt}, \\
&V_{JJ} = \frac{\hbar}{2e} \frac{d\varphi}{dt},
\end{aligned}
\end{equation}
where $R$ and $C$ are the junction resistance and capacitance. We input different current-phase relations for $I_{JJ} (\varphi)$ for different order parameter symmetries. Specifically, here we consider $\sin(3\varphi)$ case for the chiral order parameter and mixed first and second order harmonics case for the nodal order parameter. 
By solving Eq. 40 numerically, we derive the junction dynamic behavior. As shown in Figs. \ref{figure8}a and b, the $I_{JJ} (\varphi) = I_{c3} \sin(3\varphi)$ relation is reflected as fractional steps as $V_{dc} = \frac{n}{3}\frac{\hbar \omega_ac}{2e}$. For the mixed case $I_{JJ}(\varphi) = I_{c1} \sin(\varphi)+I_{c2} \sin(2\varphi)$ in Figs. \ref{figure8}c and d, we see half integer steps as $V_{dc} = \frac{n}{2}\frac{\hbar \omega_ac}{2e}$.
Here, we used the following parameters: microwave frequency $\sim 6.4$GHz, critical current $\sim 5\mathrm{nA}/\mu \mathrm{m}^2$, normal state resistance $\sim 5\mathrm{k}\Omega /\mu \mathrm{m}^2$ and interlayer geometrical capacitance $20 \mathrm{fF}/ \mu \mathrm{m}^2$. The geometric capacitance is estimated from the interlayer distance $d \sim 2\mathrm{nm}$. 

\subsection{Josephson Diode effect}
\begin{figure}[h]
    \centering
    \includegraphics[width=\linewidth]{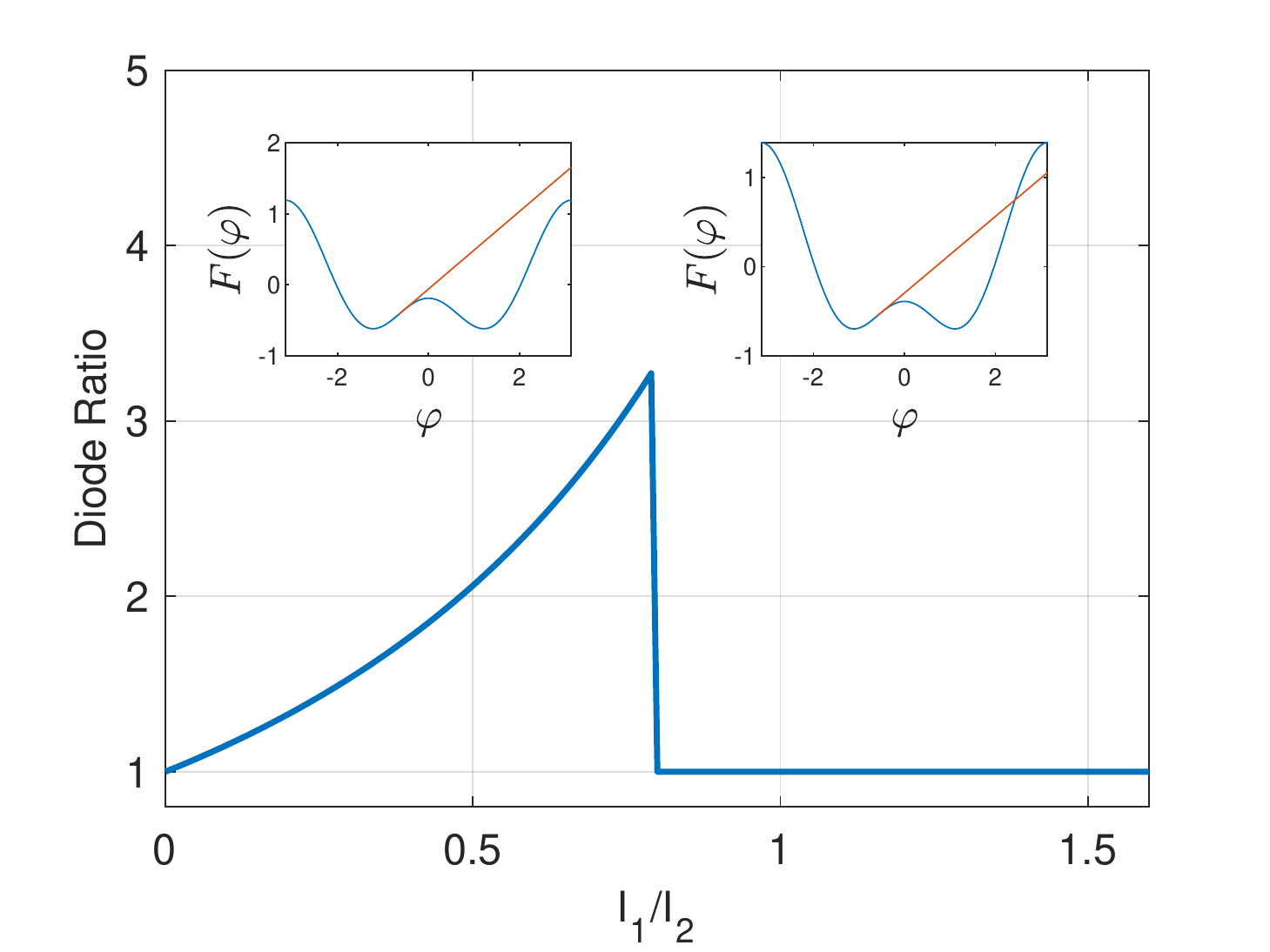}
    \caption{\textbf{Diode ratio for a junction with spontaneously broken TRS.} 
    We define the diode ratio as the ratio between the larger to the smaller critical current through the junction. Plotted here is the diode ratio as a function of the ratio between the first and second harmonic coefficients ($F=-I_1\cos(\varphi)+(I_2/2)\cos(2\varphi)$). Its maximal value is $3.27$, obtained for $I_1/I_2 \approx 0.8$ (a larger diode can be achieved by including higher order terms.). For larger $I_1/I_2$, the diode ratio is $1$, since once the current exceeds the lower of the two critical currents near one of the minima and the phase is re-trapped at the other minimum of $F(\varphi)$. The top-left inset shows the $F(\varphi)$ as a function of relative phase for $I_1/I_2<0.8$ which results in a diode effect. The line in red is tangent to the free energy curve which corresponds to the smaller of the two critical currents. The top-right is for $0.8<I_1/I_2<2$ where time-reversal is broken but there is no diode effect.}
    \label{figure9}
\end{figure}
In cases where time reversal symmetry is spontaneously broken, the critical current through the junction can depend on the direction of the current \cite{jiang2022superconducting}. 
We can define a measure of the asymmetry as the ratio between the two critical currents. 
An intuitive argument can be made regarding the possibility to have a diode effect using the washboard potential picture. 
Assuming that time reversal symmetry is broken, the ground state has a phase difference $\varphi$ which, in general, is neither $0$ nor $\pi$. 
Around $\varphi$, the phase-dependent free energy, $F(\varphi)$, is not symmetric.  
There are two points of maximal (minimal) slope, which determine the externally applied current required to drive the system out of the local minimum. 
Once the critical current is exceeded, the shape of $F(\varphi)$ could be such that the phase is re-trapped in the other minimum (Fig. \ref{figure9}, top right), or the phase keeps increasing indefinitely under the influence of the dc current, corresponding to a dissipative state (Fig. \ref{figure9}, top left). 
% If the $E-\phi$ relations are such that driving the system out of the local minima result in infinite sliding of the relative phase along the washboard potential (fig. 9 top-left), these local maximal(minimal) slopes correspond to the positive (negative) critical currents. And they are generically different values. The other case is where instead of infinite sliding, by going over the maximal slope the system just move from one minima to the other (fig. 9 top-right). 
In the case where re-trapping occurs, since the total $F(\varphi)$ relation is time reversal symmetric, the critical currents in the two directions are equal.
% we will find that the critical current (the current over which there is an infinite slide of the phase on the washboard potential) is symmetric even though TRS is broken.  

\bibliography{References}

\appendix
\onecolumngrid

\end{document}